\documentclass[11pt, reqno]{amsart}
\usepackage{amsfonts,latexsym}
\usepackage{amsmath}
\usepackage{amscd}
\usepackage{float,amsmath,amssymb,mathrsfs,bm,multirow,graphics}
\usepackage[dvips]{graphicx}
\usepackage[percent]{overpic}
\usepackage[numbers,sort&compress]{natbib}

\newcommand{\nequation}{\setcounter{equation}{0}}

\newcommand{\R}{{\Bbb R}}

\newcommand{\C}{{\Bbb C}}

\newcommand{\proofbegin}{\noindent{\it Proof.\,\,}}
\newcommand{\proofend}{\hfill$\Box$\bigskip}

\DeclareMathOperator{\im}{Im}

\newcommand{\lot}{\text{\upshape lower order terms}}

\newcommand{\res}{\text{\upshape Res\,}}

\allowdisplaybreaks

\newtheorem{theorem}{Theorem}[section]

\newtheorem{lemma}[theorem]{Lemma}

\newtheorem{figuretext}{Figure}

\input epsf
\title[The KdV equation on the half-line]{\sc The KdV equation on the half-line: The Dirichlet to Neumann map}

\author{Jonatan Lenells}
\address{Department of Mathematics, Baylor University, One Bear Place \#97328, Waco, TX 76798, USA, and Department of Applied Mathematics and Theoretical Physics, University of Cambridge, Cambridge CB3 0WA, UK.}
\email{Jonatan\_Lenells@baylor.edu}

\begin{document}

\begin{abstract} 
\noindent
We consider initial-boundary value problems for the KdV equation $u_t + u_x + 6uu_x + u_{xxx} = 0$ on the half-line $x \geq 0$. For a well-posed problem, the initial data $u(x,0)$ as well as one of the three boundary values $\{u(0,t), u_x(0,t), u_{xx}(0,t)\}$ can be prescribed; the other two boundary values remain unknown. We provide a characterization of the unknown boundary values for the Dirichlet as well as the two Neumann problems in terms of a system of nonlinear integral equations. The characterizations are effective in the sense that the integral equations can be solved perturbatively to all orders in a well-defined recursive scheme.
\end{abstract}

\maketitle



%\tableofcontents

\section{Introduction}\nequation
The main difficulty when analyzing initial-boundary value (IBV) problems for integrable PDEs is that  only a subset of the boundary values can be prescribed for a well-posed problem---the remaining boundary values are initially unknown and must be determined as part of the solution. The characterization of the unknown boundary values in terms of the prescribed data is referred to as the (generalized) Dirichlet to Neumann map. Here we analyze the Dirichlet to Neumann map for the Korteweg-de Vries (KdV) equation posed on the positive half-line:
\begin{align}\label{kdv}
  u_t + u_x + 6uu_x + u_{xxx} = 0, \qquad x > 0, \quad t > 0. 
\end{align}
For the Dirichlet problem, the initial data $u(x,0)$ as well as the boundary data $u(0,t)$ are prescribed, whereas the Neumann data $u_x(0,t)$ and $u_{xx}(0,t)$ are initially unknown. Similarly, for the first (resp. second) Neumann problem, $u(x,0)$ and $u_x(0,t)$ (resp. $u_{xx}(0,t)$) are prescribed, whereas the boundary values $u(0,t)$ and $u_{xx}(0,t)$ (resp. $u_x(0,t)$) are unknown. 
By analyzing the so-called global relation associated with the IBV problem (\ref{kdv}), we present a characterization of the unknown boundary values for the Dirichlet as well as the two Neumann problems in terms of a system of nonlinear integral equations. The characterizations are effective in the sense that the integral equations can be solved perturbatively to all orders in a well-defined and constructive recursive scheme.

The well-posedness of the IBV problem (\ref{kdv}) was analyzed in \cite{BSZ2002, BW1983} using methods of functional analysis. It was shown in \cite{BSZ2002} that the Dirichlet problem is locally well-posed for initial data in $H^s(\R)$ and boundary data in $H^{\frac{s+1}{3}}(\R^+)$ provided that $s > 3/4$. More recently, well-posedness results requiring even lower degrees of regularity have been obtained in weighted Sobolev spaces \cite{BSZ2008}. 

Our approach is based on the integrability of (\ref{kdv}) and utilizes ideas from the unified transform methodology introduced by Fokas in \cite{F1997}. Within this framework, the problem of solving the KdV equation on the half-line was first considered in \cite{F2002}, while the Dirichlet to Neumann map of (\ref{kdv}) was studied in \cite{TF2008}, where integral equations characterizing the unknown Neumann values $\{u_x(0,t), u_{xx}(0,t)\}$ for the Dirichlet problem were derived using the so-called Gelfand-Levitan-Marchenko (GLM) representations of the eigenfunctions of the associated Lax pair. The approach of \cite{TF2008} was first developed for the nonlinear Schr\"odinger (NLS) equation in \cite{BFS2003} and subsequently implemented also for the modified KdV and sine-Gordon equations \cite{F2005}. 
%Although the derivations in \cite{F2005} were based on the GLM representations, the final formulas did not involve these representations.
Recently, in \cite{trilogy1}, a more direct approach to the Dirichlet to Neumann map was presented in which the derivation takes place entirely in the spectral space, avoiding in particular the need for any GLM representations. This approach was implemented for the NLS equation on an interval in \cite{trilogy3} and for an equation with a $3\times 3$ Lax pair in \cite{L3x3} (an early version of the approach was implemented for the derivative NLS equation on the half-line in \cite{LdnlsD2N}). 

The purpose of the present paper is to employ the ideas of \cite{trilogy1} to analyze the IBV problem (\ref{kdv}). In particular, we derive new characterizations of the Dirichlet to Neumann map for the Dirichlet and Neumann problems for (\ref{kdv}) (for the Dirichlet problem, an alternative characterization was already obtained in \cite{TF2008}). Compared with the investigations of \cite{trilogy1, LdnlsD2N, L3x3, trilogy3}, the analysis of (\ref{kdv}) is complicated by the fact that the Lax pair of (\ref{kdv}) is singular at $k = 0$, and also by the fact that the dispersion relation entering the Lax pair is not a simple power, $\omega(k) = k^n$, but has the form $\omega(k) = k-4k^3$. In the case of a simple power, certain formulas simplify since $k^n$ is invariant under the rotations $k \to e^{\frac{2\pi i}{n}} k$. The case of $k-4k^3$ is more involved, but represents a more generic situation. 

The KdV equation first appeared in work by Boussinesq and Korteweg and de Vries as a model for waves of small amplitude propagating on the surface of shallow water. 
In the context of wave propagation, IBV problems for the KdV equation arise naturally, the boundary data typically being obtained from measurements of an incoming wave shape at a fixed point in space. Examples of situations where the IBV problem (\ref{kdv}) is relevant include the modeling of near-shore wave motion generated by waves propagating from deep water, as well as the generation of waves in laboratory experiments where a wave maker is mounted at one end of a wave tank \cite{BPS1981, BSZ2002}.

Finally, we mention that the KdV equation (\ref{kdv}) is often brought to the form $u_t + 6uu_x + u_{xxx} = 0$ by means of a Galilean transformation. If the $u_x$-term is removed from (\ref{kdv}), the dispersion relation reduces to $\omega(k) = -4k^3$ and the analysis below simplifies accordingly. However, we emphasize that such a Galilean transformation when applied to (\ref{kdv}) turns the half-line problem into a problem with moving boundary, which is typically not the situation relevant for applications. Therefore, we choose to keep the $u_x$-term in (\ref{kdv}).

\section{Lax pair and eigenfunctions}\nequation
Let $u(x,t)$ be a real-valued solution of the KdV equation (\ref{kdv}) in the half-line domain
$$\Omega = \{(x,t) \in \R^2 \, | \, 0 < x < \infty \;\; \text{and} \;\; 0 < t < T\},$$
where $0 < T < \infty$ denotes a given final time.
Let $\{g_j(t)\}_0^2$ denote the boundary values of $u(x,t)$: 
$$g_0(t) = u(0,t), \qquad g_1(t) = u_x(0,t), \qquad g_2(t) = u_{xx}(0,t).$$
Equation (\ref{kdv}) admits the Lax pair
\begin{align}\label{lax}
\begin{cases}
\mu_x + ik[\sigma_3, \mu] = V_1 \mu,
	\\
\mu_t + i(4k^3 - k)[\sigma_3, \mu] = V_2 \mu,
\end{cases}
\end{align}
where $\mu(x,t,k)$ is a $2\times2$-matrix valued eigenfunction, $k \in \C$ is the spectral parameter, the functions $V_1(x,t,k)$ and $V_2(x,t,k)$ are defined by
\begin{align*}
& V_1 = \frac{u}{2k} (\sigma_2 - i \sigma_3),
	\\
& V_2 = 2ku\sigma_2 + u_x \sigma_1 + \frac{2u^2 + u + u_{xx}}{2k}(i\sigma_3 - \sigma_2),
\end{align*}
and $\{\sigma_j\}_1^3$ denote the standard Pauli matrices.
\begin{figure}
\begin{center}
\begin{overpic}[width=.5\textwidth]{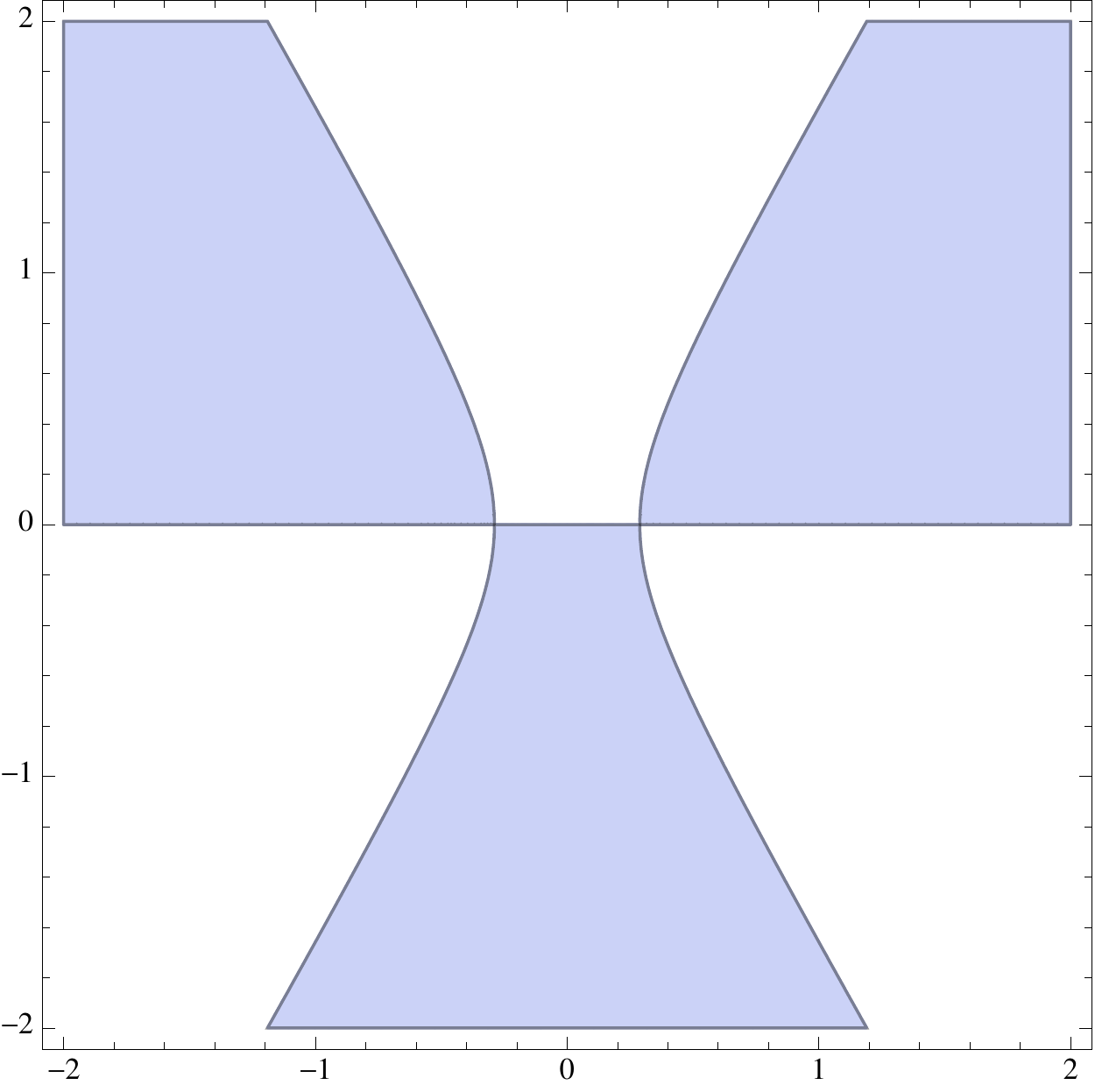}
      \put(77,67){$D_1'$}
      \put(48,75){$D_2$}
      \put(17,67){$D_1''$}
      \put(17,30){$D_4''$}
      \put(48,22){$D_3$}
      \put(77,30){$D_4'$}
%       \put(105,50){$\text{Re}\; k$}
%      \put(60,47){$\pi/3$}
      \end{overpic}
     \begin{figuretext}\label{DjsKdV.pdf}
       The domains $\{D_j\}_1^4$ in the complex $k$-plane with $D_1 = D_1' \cup D_1''$ and $D_4 = D_4' \cup D_4''$. 
     \end{figuretext}
     \end{center}
\end{figure}
We define three eigenfunctions $\{\mu_j(x,t,k)\}_1^3$ of (\ref{lax}) as the solutions of the Volterra integral equations
\begin{align}\label{mujdef}
  \mu_j(x,t,k) = I + \int_{(x_j, t_j)}^{(x,t)} e^{-ik(x - x')\hat{\sigma}_3 - i(4k^3-k)(t - t')\hat{\sigma}_3} W_j(x',t',k), \qquad
  (x,t) \in \Omega,
\end{align}
where $W_j = (V_1 dx + V_2 dt)\mu_j$, $(x_1,t_1) = (0,T)$, $(x_2,t_2) =(0,0)$, $(x_3,t_3) = (\infty, t)$, and $\hat{\sigma}_3$ acts on a $2\times 2$ matrix $A$ by $\hat{\sigma}_3A = [\sigma_3, A]$, i.e. $e^{\hat{\sigma}_3} A = e^{\sigma_3} A e^{-\sigma_3}$.
These functions satisfy the symmetry
\begin{align}\label{mujsymm}
\mu_j(x,t,k) = \sigma_1 \overline{\mu_j(x,t,\bar{k})} \sigma_1, \qquad j = 1,2,3.
\end{align}
We define the spectral functions $a(k)$ and $b(k)$ by
\begin{align}\label{sSdef}  
  \begin{pmatrix} \overline{a(\bar{k})} & b(k) \\
  \overline{b(\bar{k})} & a(k) \end{pmatrix}
   = \mu_3(0,0,k),
 % \qquad \begin{pmatrix} \overline{A(\bar{k})} & B(k) \\  \overline{B(\bar{k})} & A(k) \end{pmatrix} = \mu_1(0,0,k).
\end{align}
and the open subsets $\{D_j\}_1^4$ of the complex $k$-plane by 
\begin{align*}
D_1 = \{\im k > 0 \cap \im(4k^3 - k) > 0\},  \qquad
D_2 = \{\im k > 0 \cap \im(4k^3 - k) < 0\}, 
	\\
D_3 = \{\im k < 0 \cap \im(4k^3 - k) > 0\},  \qquad
D_4 = \{\im k < 0 \cap \im(4k^3 - k) < 0\}.
\end{align*}	
Let $D_1 = D_1' \cup D_1''$ where $D_1' = D_1 \cap \{\text{Re}\, k > 0\}$ and $D_1'' = D_1 \cap \{\text{Re}\, k < 0\}$. Similarly, let $D_4 = D_4' \cup D_4''$ with $D_4' = D_4 \cap \{\text{Re}\, k > 0\}$ and $D_4'' = D_4 \cap \{\text{Re}\, k < 0\}$, see Figure \ref{DjsKdV.pdf}.
The functions $a(k)$ and $b(k)$ are analytic and bounded in $D_1 \cup D_2$ except for a possible singularity at $k = 0$. 
%while $A(k)$ and $B(k)$ are entire functions which are bounded in $D_1 \cup D_3$ except for a possible singularity at $k = 0$. 
%The solution $u(x,t)$ can be reconstructed from the solution of a matrix Riemann-Hilbert problem with jump matrix defined in terms of $\{a(k), b(k), A(k), B(k)\}$, see \cite{F2002}. However, since $\{A(k), B(k)\}$ are defined in terms of both the Dirichlet and the Neumann data, an effective solution of the problem requires a characterization of the Dirichlet to Neumann map. 

%Differential equations for \Phi_1 and \Phi_2:
%\begin{align*}
%  \Phi_{1t} + 2i(4k^3 - k)\Phi_1 =  \frac{i}{2k}(u + 2u^2 + u_{xx}) \Phi_1 + (-2ik u + u_x + \frac{i}{2k}(u + 2u^2 + u_{xx}))\Phi_2,
%  	\\
%\Phi_{2t} = (2iku + u_x - \frac{i}{2k}(u + 2u^2 + u_{xx})) \Phi_1 - \frac{i}{2k}( u + 2u^2 + u_{xx})\Phi_2.	
%\end{align*}

\subsection{The global relation}
Let $\Phi_1(t,k)$ and $\Phi_2(t,k)$ denote the $(12)$ and $(22)$ entries of $\mu_2(0,t,k)$.
Then $\Phi_1(t, k)$ and $\Phi_2(t, k)$ are analytic in $\C \setminus \{0\}$ and bounded as $k \to \infty$ inÊ $\bar{D}_2 \cup \bar{D}_4$. Moreover, the spectral functions $\{\Phi_j\}_1^2$ and $\{a(k), b(k)\}$ satisfy an important global relation. More precisely, define $R(t,k)$ by
$$R(t, k) = \frac{b(k)\overline{\Phi_2(t, \bar{k})}}{a(k)}, \qquad 0 < t < T, \quad \im k \geq 0.$$
Then the function $c(t,k)$ defined by
\begin{align}\label{GRc}
& c(t, k) = \Phi_1(t, k) + R(t, k) e^{-2i(4k^3-k)t}, \qquad 0 < t < T, \quad \im k \geq 0,
\end{align} 
satisfies
\begin{equation}\label{cddef}
  c(t, k) = \frac{F(t,k)}{a(k)},
\end{equation}
where $F(t,k)$ is a function which is analytic in $\im k > 0$, continuous in $\{\im k \geq 0, k\neq 0\}$, and $F(t,k) = O(k^{-2})$ as $k \to \infty$ in $\im k \geq 0$ cf. \cite{trilogy1}.
We will use these properties of $c(t,k)$ to characterize the Dirichlet to Neumann map.

\subsection{Asymptotics as $k \to \infty$}
The eigenfunctions $\{\mu_j\}_1^3$ admit the asymptotics
\begin{subequations}\label{mujasymptotics}
\begin{align}\nonumber
 (\mu_j(x,t,k))_{12} = &\; \frac{\mu_{j12}^{(2)}(x,t) }{k^2} + \frac{\mu_{j12}^{(3)}(x,t) }{k^3}
+ \frac{\mu_{j12}^{(4)}(x,t) }{k^4} + O\Bigl(\frac{1}{k^5}\Bigr) 
	\\ \label{mujasymptoticsa}
& + \begin{cases}
O\Bigl(\frac{e^{-2ik(x - x_j)-2i(4k^3-k)(t-t_j)}}{k^2}\Bigr), & j =1,2,
	\\
0, & j = 3.
\end{cases}	 \qquad k \to \infty,
	\\ \label{mujasymptoticsb}
(\mu_j(x,t,k))_{22} = &\; 1 + \frac{\mu_{j22}^{(1)}(x,t)}{k} + \frac{\mu_{j22}^{(2)}(x,t)}{k^2} + O\Bigl(\frac{1}{k^3}\Bigr), \qquad k \to \infty, %???Need exponential term? I think it is of O(k^{-3})
\end{align}
\end{subequations}
where
\begin{subequations}
\begin{align} \label{muj12coefficients}
& \mu_{j12}^{(2)}% = \mu_{j21}^{(2)} 
= -\frac{u}{4}, \quad
\mu_{j12}^{(3)}% = -\mu_{j21}^{(3)} 
= \frac{u_x}{8i} - \frac{u}{4}\mu_{j22}^{(1)},
\quad \mu_{j12}^{(4)} %= \mu_{j21}^{(4)}
 = \frac{u^2}{8} + \frac{u_{xx}}{16} - \frac{u}{4}\mu_{j22}^{(2)} + \frac{u_x}{8i} \mu_{j22}^{(1)},
	\\ \label{muj22coefficients}
& \mu_{j22}^{(1)} %= -\mu_{j11}^{(1)} 
= \frac{i}{2}\int_{(x_j, t_j)}^{(x,t)} \Delta, \qquad 
\mu_{j22}^{(2)} %= \mu_{j11}^{(2)} 
= -\frac{1}{8}\left(\int_{(x_j,t_t)}^{(x,t)} \Delta\right)^2,
%	\\ \nonumber
%& \mu_{j22}^{(3)} = -\mu_{j11}^{(3)} = \int_{(x_j, t_j)}^{(x,t)} \bigg\{\bigg(\frac{iu}{2}\mu_{j22}^{(2)} - \frac{iu^2}{8}\bigg)dx 
%	\\
%& + (2iu\mu_{j12}^{(4)} + u_x \mu_{j12}^{(3)} - \frac{i}{2}(u + 2u^2 + u_{xx})\mu_{j12}^{(2)} - \frac{i}{2}(u + 2u^2 + u_{xx})\mu_{j22}^{(2)})dt \bigg\},	
\end{align}
\end{subequations}
 the closed one-form $\Delta$ is defined by
\begin{align*}
 \Delta & = u dx - (u + 3u^2 + u_{xx})dt,
\end{align*}
and the expansions in (\ref{mujasymptotics}) are valid for $k$ approaching $\infty$ within the regions of boundedness of $(\mu_j)_{12}$ and $(\mu_j)_{22}$.
Indeed, integration by parts in the $(12)$ entry of the Volterra integral equation (\ref{mujdef}) yields (\ref{mujasymptoticsa}) with the $\mu_{j12}^{(n)}$'s given by (\ref{muj12coefficients}). Substitution of this expansion into the $(22)$ entry of (\ref{mujdef}) shows that (\ref{mujasymptoticsb}) holds with the $\mu_{j22}^{(n)}$'s given by (\ref{muj22coefficients}).

Equations (\ref{mujasymptotics}) imply that $\{\Phi_j\}_1^2$ admit the following asymptotics as $k \to \infty$, $k \in \bar{D}_2 \cup \bar{D}_4$:
\begin{subequations} \label{Phiabcexpansions}
\begin{align} \label{Phiabcexpansionsa}
& \Phi_1(t, k) = \frac{\Phi_1^{(2)}(t)}{k^2} + \frac{\Phi_1^{(3)}(t)}{k^3} + \frac{\Phi_1^{(4)}(t)}{k^4} + O \Bigl(\frac{1}{k^5} \Bigr) + O\Bigl(\frac{e^{-2i(4k^3-k)t}}{k^2}\Bigr), 
	\\ \label{Phiabcexpansionsb}
& \Phi_2(t, k) = 1 + \frac{\Phi_2^{(1)}(t)}{k} + \frac{\Phi_2^{(2)}(t)}{k^2}  + O \Bigl(\frac{1}{k^3} \Bigr), 
\end{align}
\end{subequations}
where 
\begin{align*}
& 
\Phi_2^{(1)} = \frac{i}{2} \int_{(0, 0)}^{(0,t)} \Delta, \qquad
\Phi_2^{(2)} = -\frac{1}{8}\left(\int_{(0,0)}^{(0,t)} \Delta\right)^2,
	\\
& \Phi_1^{(2)} = -\frac{g_0}{4}, \qquad \Phi_1^{(3)} = \frac{g_1}{8i} - \frac{g_0}{4}\Phi_{2}^{(1)},
\qquad
 \Phi_1^{(4)} = \frac{g_0^2}{8} + \frac{g_2}{16} - \frac{g_0}{4}\Phi_2^{(2)} + \frac{g_1}{8i} \Phi_2^{(1)}.	
\end{align*}
In particular, we find the following expressions for the boundary values:
\begin{subequations}\label{g0g1g2}
\begin{align}\label{g0Phi}
& g_0 = -4\Phi_1^{(2)},
    	\\\label{g1Phi}
& g_1 = 2ig_0\Phi_2^{(1)} + 8i \Phi_1^{(3)},
	\\\label{g2Phi}
& g_2 = 16\Phi_{1}^{(4)} - 2g_0^2 + 4g_0\Phi_{2}^{(2)} + 2ig_1\Phi_{2}^{(1)}.
\end{align}
\end{subequations}
For $j = 3$, the equations in (\ref{mujasymptotics}) are valid for $\im k \geq 0$ and evaluation at $x=t=0$ gives
\begin{align*}
  a(k) = 1 + \frac{\mu_{322}^{(1)}(0,0)}{k} + O\Bigl(\frac{1}{k^2}\Bigr), \qquad b(k) = O\Bigl(\frac{1}{k^2}\Bigr),  \qquad k \to \infty, \quad \im k \geq 0.
\end{align*}

We will also need the following result describing the asymptotics of the function $c(t,k)$ defined in (\ref{cddef}).

\begin{lemma}\label{clemma}
The global relation (\ref{GRc}) implies that
\begin{align} \label{casymptotics}
& c(t, k) = \frac{\Phi_1^{(2)}(t)}{k^2} + \frac{\Phi_1^{(3)}(t)}{k^3} + \frac{\Phi_1^{(4)}(t)}{k^4} +  O \Bigl(\frac{1}{k^5} \Bigr), \qquad k \to \infty, \quad k \in \bar{D}_1.
\end{align}
\end{lemma}
\proofbegin
The proof is similar to the proof in Appendix B of \cite{trilogy1}.
\proofend

\subsection{Asymptotics as $k \to 0$}
Since the functions $V_1, V_2$ have simple poles at $k = 0$, the solutions of (\ref{lax}) will, in general, be  singular at $k = 0$.

\begin{lemma}
We have
\begin{align}\label{Phiat0}
& \begin{pmatrix} \Phi_1(t,k) \\ \Phi_2(t,k) \end{pmatrix}
 = \frac{i \alpha(t)}{k}\begin{pmatrix} 1 \\ -1 \end{pmatrix} + O(1), \qquad k \to 0, \quad k \in \C,
\end{align}
where $\alpha(t)$ is a real-valued function. Moreover, 
\begin{align} \label{abat0}
& \begin{pmatrix} b(k) \\ a(k) \end{pmatrix}
 = \frac{i \beta}{k}\begin{pmatrix} 1 \\ -1 \end{pmatrix} + O(1), \qquad k \to 0, \quad \im k \geq 0,
\end{align}  
where $\beta \in \R$ is a constant. 
\end{lemma}
\proofbegin
The second column of the $x$-part of the Lax pair (\ref{lax}) is
$$\begin{cases}
\mu_{12x} + 2ik\mu_{12} = -\frac{iu}{2k}(\mu_{12} + \mu_{22}),
	\\
\mu_{22x} = \frac{iu}{2k}(\mu_{12} + \mu_{22}).
\end{cases}$$
It follows that the function $f = (\mu_3)_{12} + (\mu_3)_{22}$ satisfies
$$f_{xx} + 2ikf_x + uf = 0$$
and $f \sim 1$ as $x \to \infty$, i.e. $f$ is a Jost function. Standard Sturm-Liouville theory (see Lemma 1 in \cite{DT1979}) implies that $f$ and $\dot{f} := df/dk$ are continuous in $\im k \geq 0$ for any $(x,t) \in \Omega$ and that
\begin{align}\label{fxtkat0}
f(x,t,k) = f(x,t,0) + \dot{f}(x,t,0) k + O(k^2), \qquad k \to 0, \quad \im k \geq 0.
\end{align}
The reality condition $f(x,t,k) = \overline{f(x,t,-\bar{k})}$ implies that $f(x,t,0)\in \R$  whereas  $\dot{f}(x,t,0) \in i\R$.
Since, for $\im k \geq 0$,
\begin{align*}
(\mu_{3}(x,t,k))_{12}  = &\; \frac{i}{2k} \int_x^\infty e^{2ik(x'-x)} (u f)(x',t,k) dx'
%	\\
%& = \frac{i}{2k} \int_x^\infty u f dx' + \int_x^\infty e^{2ik(x'-x)} \bigg(\int_\infty^{x'} (uf)(x'',t,k)dx''\bigg) dx'
	\\
= &\; \frac{i}{2k} \int_x^\infty (u f)(x',t,k) dx' - \int_x^\infty \int_{x'}^\infty(uf)(x'', t,k) dx'' dx'
	\\
&-  2ik \int_x^\infty e^{2ik(x'-x)} \bigg(\int_{x'}^\infty \int_{x''}^\infty (uf)(x''',t,k)dx'''dx''\bigg) dx'
\end{align*}
and
\begin{align*}
(\mu_{3}(x,t,k))_{22} & = -\frac{i}{2k} \int_x^\infty (u f)(x',t,k) dx',
\end{align*}
equation (\ref{fxtkat0}) and the symmetry (\ref{mujsymm}) yield
\begin{align}\nonumber
& \mu_3(x,t,k) = \frac{i \alpha_3(x,t)}{k}\begin{pmatrix} 1 & 1 \\ -1 & -1 \end{pmatrix} + \begin{pmatrix} \beta_3(x,t) & \gamma_3(x,t) \\
\gamma_3(x,t) & \beta_3(x,t) \end{pmatrix} + O(k), 
	\\ \label{mu3at0}
&\hspace{7cm} k \to 0, \quad \im k \geq 0,
\end{align}
where $\alpha_3(x,t) = \frac{1}{2} \int_x^\infty (uf)(x',t,0) dx'$ and $\beta_3, \gamma_3$ are real-valued functions.
Evaluating (\ref{mu3at0}) at $x=t=0$, we obtain (\ref{abat0}). 
Using (\ref{mu3at0}) in the relation
$$\mu_2 = \mu_3 e^{-i(kx + (4k^3-k)t)\hat{\sigma}_3} \begin{pmatrix} a(k)  & -b(k) \\
  - \overline{b(\bar{k})} & \overline{a(\bar{k})}\end{pmatrix}$$
evaluated at $x = 0$, we find (\ref{Phiat0}).
\proofend

\section{The generalized Dirichlet to Neumann map}\nequation
We can now derive an effective characterization of the generalized Dirichlet to Neumann map for the Dirichlet ($g_0$ prescribed), the first Neumann ($g_1$ prescribed), and the second Neumann ($g_2$ prescribed) problems. The derivation relies heavily on the invariance properties of the dispersion relation $\omega(k) = k-4k^3$.

For each $k \in \C$, let $\nu_j = \nu_j(k)$, $j = 1,2,3$, denote the three roots of the following cubic equation in $\nu$:
$$4\nu^3 - \nu - (4k^3 - k) = 0.$$
It is not possible to choose a consistent numbering of the $\nu_j(k)$'s as $k$ varies over the whole complex plane. Indeed, one of the $\nu_j(k)$'s equals $k$ and the two other roots are given by $-\frac{1}{2}(k \pm \sqrt{1 - 3k^2})$. Hence, as $k$ encircles one of the points $k = \pm 1/\sqrt{3}$, the other two roots are interchanged. However, it is possible to fix a numbering of the roots $\{\nu_j(k)\}_1^3$ for $k$ lying in the restricted set $\bar{D}_1 \cup \bar{D}_3$. For $k \in \bar{D}_1 \cup \bar{D}_3$, each of the three sets $\bar{D}_1'$, $\bar{D}_1''$, and $\bar{D}_3$ contains exactly one root\footnote{If $\nu_1(k) = 1/\sqrt{3}$, then 
$$\nu_2(k) = \nu_3(k) =  - \frac{1}{2\sqrt{3}} \in \bar{D}_1'' \cap \bar{D}_3$$ 
is a double root, whereas if $\nu_2(k) = -1/\sqrt{3}$, then 
$$\nu_1(k) = \nu_3(k) = \frac{1}{2\sqrt{3}} \in \bar{D}_1' \cap \bar{D}_3$$ 
is a double root.}; we denote these roots by $\nu_1(k)$, $\nu_2(k)$, and Ê$\nu_3(k)$ respectively. 
%As $k \to 0$,
%$$\nu_1(k) = \frac{1}{2} - \frac{k}{2} - \frac{3k^2}{4} + O(k^3), \qquad \nu_2(k) = -\frac{1}{2} - \frac{k}{2} + \frac{3k^2}{4} + O(k^3), \qquad k \to 0.$$
Identification of the coefficients of $\nu^n$, $n = 0,1,2$, in 
\begin{align}\label{nuzeros}
4\nu^3 - \nu - (4k^3 - k) = 4(\nu - \nu_1(k))(\nu - \nu_2(k))(\nu - \nu_3(k))
\end{align}
yields the identities
\begin{align} \nonumber
& \nu_1(k)\nu_2(k)\nu_3(k) = \frac{4k^3 - k}{4}, \qquad \nu_1(k)\nu_2(k) + \nu_2(k)\nu_3(k) + \nu_3(k) \nu_1(k) = -\frac{1}{4},
	\\ \label{nurelations}
&  \nu_1(k) + \nu_2(k) + \nu_3(k) = 0.	
\end{align}

\begin{figure}
\begin{center}
\begin{overpic}[width=.4\textwidth]{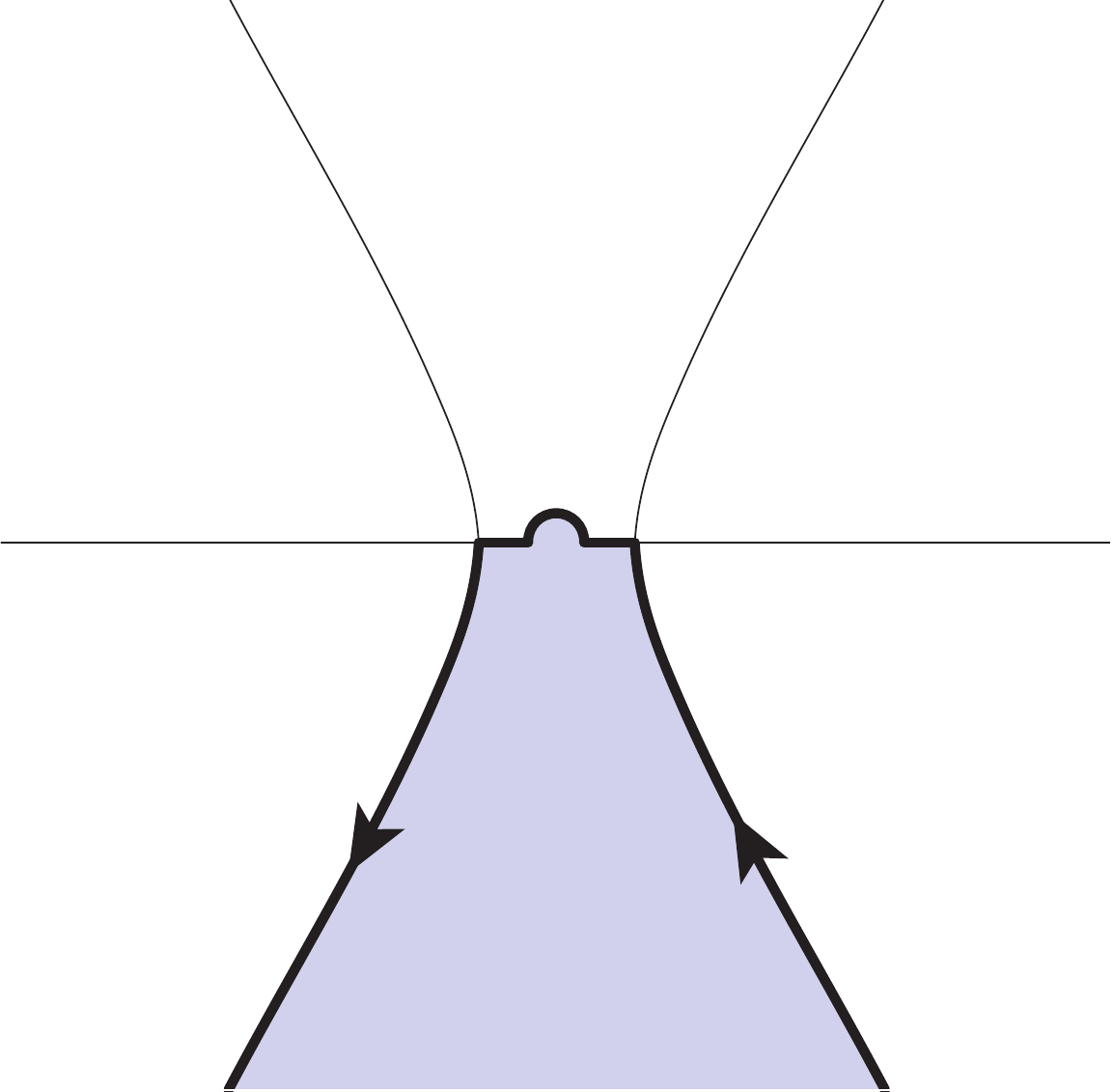}
       \put(71,23){$\partial \hat{D}_3$}
      \put(47,12){$\hat{D}_3$}
      \end{overpic}
      \qquad\quad
      \begin{overpic}[width=.4\textwidth]{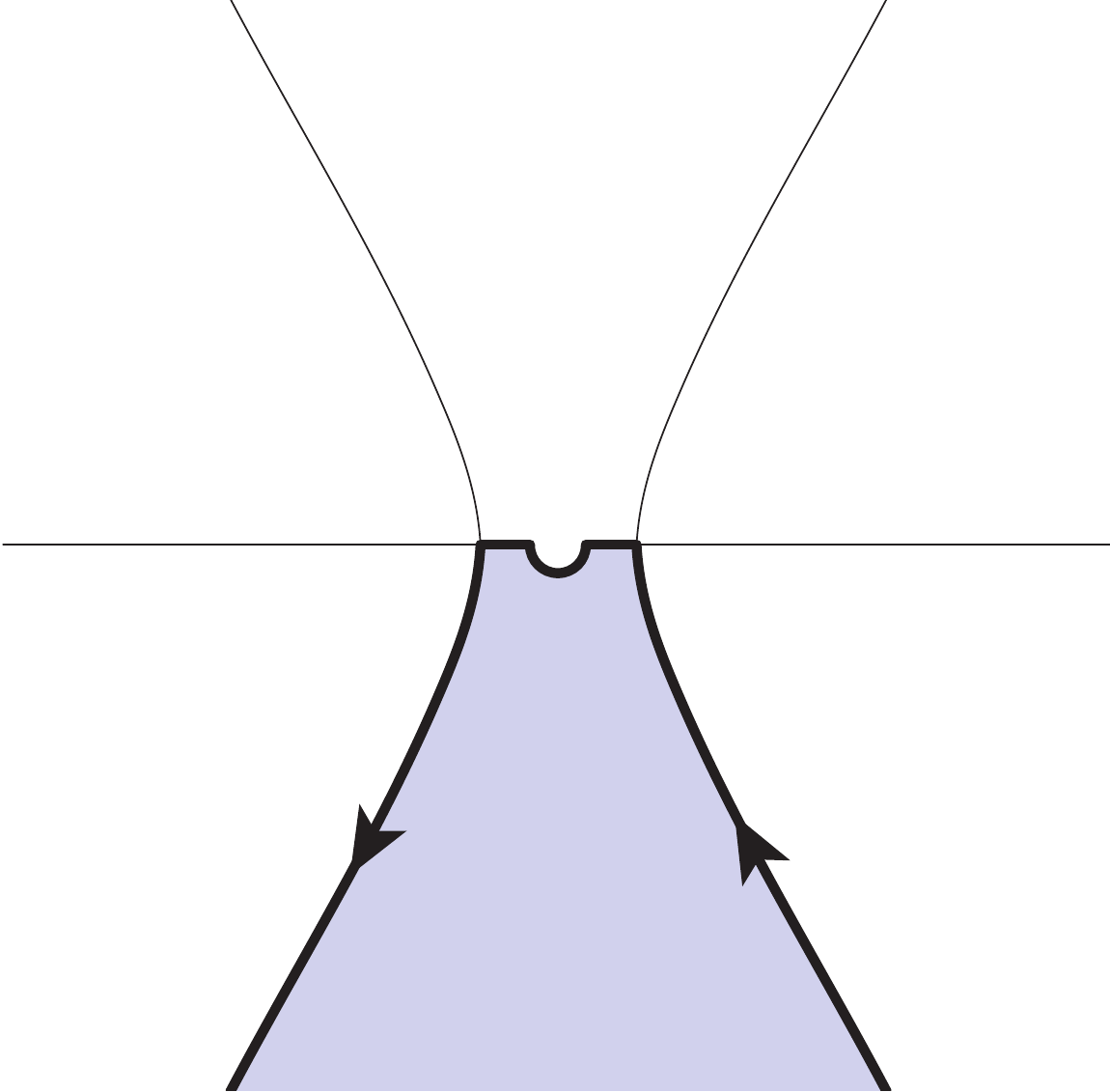}
      \put(71,23){$\partial \check{D}_3$}
      \put(47,12){$\check{D}_3$}
      \end{overpic}
     \begin{figuretext}\label{D3hatcheckfig}
       The contours $\partial \hat{D}_3$ and $\partial \check{D}_3$.
     \end{figuretext}
     \end{center}
\end{figure}

Since $\{\Phi_1, \Phi_2\}$ have simple poles at $k = 0$, we will occasionally need to deform the integration contours so that they pass around the points $k = 0$, $k = \nu_1(0) = \frac{1}{2}$, and $k = \nu_2(0) = -\frac{1}{2}$. Thus, let $\partial \hat{D}_3$ and $\partial \check{D}_3$ denote the contour $\partial D_3$ with a small indentation inserted at $k = 0$ such that the indentation lies in $D_2$ for $\partial \hat{D}_3$ and in $D_3$ for $\partial \check{D}_3$, see Figure \ref{D3hatcheckfig}. More generally, if either of the points $\{0, \pm \frac{1}{2}\}$ lies on a contour $\gamma$, then $\hat{\gamma}$ and $\check{\gamma}$ will denote the contour $\gamma$ with small indentations inserted at the points $\{0, \pm \frac{1}{2}\}$ such that the indentations lie in $D_2 \cup D_4$ for $\hat{\gamma}$ and in $D_1 \cup D_3$ for $\check{\gamma}$.
Note that as $k$ traverses the contour $\partial D_3$, $\nu_1(k)$ and $\nu_2(k)$ traverse the contours $\partial D_1'$ and $\partial D_1''$ respectively.

Let $\{\chi_j, \hat{\chi}_j, \check{\chi}_j, \tilde{\chi}_j\}_1^2$ denote the following symmetric combinations formed from $\{\Phi_j\}_1^2$: 
\begin{align}\nonumber
& \chi_j(t,k) = \sum_{m=1}^3 \nu_m'(k) \Phi_j(t, \nu_m(k)), \qquad j = 1,2,
	\\ \nonumber
& \hat{\chi}_j(t,k) = \sum_{m=1}^3 \nu_m(k) \nu_m'(k) \Phi_j(t, \nu_m(k)), \qquad j = 1,2,
	\\	\nonumber
& \check{\chi}_j(t,k) = \sum_{m=1}^3  \nu_m^2(k) \nu_m'(k) \Phi_j(t, \nu_m(k)), \qquad j = 1,2,
	\\ \label{chidef}
& \tilde{\chi}_j(t,k) = \sum_{m=1}^3  \nu_m^3(k)\nu_m'(k)\Phi_j(t, \nu_m(k)), \qquad j = 1,2.
\end{align}

The nonlinear integral equations presented in the following theorem characterize the generalized Dirichlet to Neumann map for the Dirichlet as well as the two Neumann problems for (\ref{kdv}). 

%The spectral functions $A(k)$ and $B(k)$ are given in terms of $\Phi_1$ and $\Phi_2$ by
%\begin{align}\label{ABexpressions}
%& A(k)  = \overline{\Phi_2(T, \bar{k})},  \qquad B(k) = -\Phi_1(T, k)e^{2i(4k^3-k)T}.
%\end{align}

\begin{theorem}\label{th1}
Let $0 < T < \infty$. Let $q_0(x)$, $x \geq 0$, be a function of Schwartz class. For the Dirichlet problem it is assumed that the function $g_0(t)$, $0 \leq t < T$, has sufficient smoothness and is compatible with $q_0(x)$ at $x=t=0$. Similarly, for the first and second Neumann problems it is assumed that the functions Ê$g_1(t)$ and $g_2(t)$, $0 \leq t < T$, have sufficient smoothness and are compatible with $q_0(x)$ at $x=t=0$, respectively.
Suppose that $a(k)$ has a finite (possibly empty) set $\{k_j\}_1^N$ of simple zeros in $D_1 \cup D_2$; assume that no zeros occur on the boundaries of $D_1$ and $D_2$. 

Then the complex-valued functions $\Phi_1(t, k)$ and $\Phi_2(t, k)$ satisfy the following system of nonlinear integral equations:
\begin{subequations}\label{Phieqs}
\begin{align} \nonumber
  \Phi_1(t, k) = & \int_0^t e^{2i(4k^3-k)(t' - t)} \bigg[\frac{i}{2k}(g_0 + 2g_0^2 + g_2) \Phi_1 
	\\ \label{Phieqsa}
&   + \Big(-2ik g_0 + g_1 + \frac{i}{2k}(g_0 + 2g_0^2 + g_2)\Big)\Phi_2\bigg] (t', k) dt',
	\\ \nonumber
   \Phi_2(t, k) = &\; 1 + \int_0^t \bigg[ \Big(2ikg_0 + g_1 - \frac{i}{2k}(g_0 + 2g_0^2 + g_2)\Big) \Phi_1
  	\\ \label{Phieqsb}
&  - \frac{i}{2k}( g_0 + 2g_0^2 + g_2)\Phi_2\bigg](t', k) dt', \qquad 0 < t < T, \;\; k \in \C, \; \; k \neq 0.
\end{align}
\end{subequations}

\begin{itemize}
\item[$(a)$] For the Dirichlet problem, the unknown Neumann boundary values $g_1(t)$ and $g_2(t)$ are given by
\begin{subequations}\label{g1g2expression}
\begin{align} \nonumber
g_1(t) = \;& \frac{2g_0(t)}{\pi}\int_{\partial \hat{D}_3} \chi_2(t,k) dk 
+ \frac{4}{\pi}\int_{\partial D_3}  \biggl[k \hat{\chi}_1(t,k) + \frac{12k^2 -1}{4k^2-1} \frac{g_0(t)}{4}\biggr] dk
 	\\\nonumber
& + 8 i \sum_{k_j \in D_1} \big(k_j - \nu_3(k_j)\big)k_j e^{-2i(4k_j^3-k_j)t} \underset{k_j}{\res}R(t, k)
 	\\\nonumber
&  - \frac{4}{ \pi } \int_{\partial D_3} e^{-2i(4k^3-k)t}  \bigl[ (\nu_1(k) - k)\nu_1(k)\nu_1'(k) R(t, \nu_1(k)) 
	\\\label{g1expression}
&  + (\nu_2(k) - k)\nu_2(k)\nu_2'(k) R(t,\nu_2(k)) \bigr] dk
\end{align}
and
\begin{align} \nonumber
g_2(t) = \;& 
\frac{8}{\pi i}\int_{\partial D_3} \biggl[k^2 \hat{\chi}_1(t,k) + k\frac{12k^2 -1}{4k^2-1} \frac{g_0(t)}{4} \biggr] dk
 	\\\nonumber
& + 16 \sum_{k_j \in D_1} \big(k_j^2 - \nu_3^2(k_j)\big)k_j e^{-2i(4k_j^3-k_j)t} \underset{k_j}{\res}R(t, k)
  	\\ \nonumber
&  - \frac{8}{\pi i} \int_{\partial D_3} e^{-2i(4k^3-k)t}  \bigl[ (\nu_1^2(k) - k^2)\nu_1(k)\nu_1'(k) R(t, \nu_1(k)) 
	\\\nonumber
&  + (\nu_2^2(k) - k^2)\nu_2(k)\nu_2'(k) R(t,\nu_2(k)) \bigr] dk
	\\ \label{g2expression}
& - 2g_0^2(t) + \frac{4 g_0(t)}{i\pi} \int_{\partial D_3} \hat{\chi}_2(t,k)dk
+ \frac{2g_1(t)}{\pi}\int_{\partial \hat{D}_3} \chi_2(t,k)dk.
\end{align}
\end{subequations}

\item[$(b)$] For the first Neumann problem, the unknown boundary values $g_0(t)$ and $g_2(t)$ are given by
\begin{subequations}\label{N1expressions}
\begin{align} \nonumber
  g_0(t) =& -\frac{2}{\pi i}\int_{\partial \check{D}_3}  \frac{1}{k}\check{\chi}_1(t,k) dk
 - 4 \sum_{k_j \in D_1} \bigg(\frac{1}{k_j} - \frac{1}{\nu_3(k_j)}\bigg)k_j^2 e^{-2i(4k_j^3-k_j)t} \underset{k_j}{\res}R(t, k)
	\\\nonumber
&+ \frac{2}{\pi i} \int_{\partial \check{D}_3} e^{-2i(4k^3-k)t}  \bigg[ \bigg(\frac{1}{\nu_1(k)} - \frac{1}{k}\bigg)\nu_1^2(k) \nu_1'(k) R(t, \nu_1(k)) 
	\\ \label{N1g0expression} 
&  + \bigg(\frac{1}{\nu_2(k)} - \frac{1}{k}\bigg)\nu_2^2(k)\nu_2'(k) R(t,\nu_2(k)) \bigg] dk
\end{align}
and
\begin{align} \nonumber
  g_2(t) =\;&
 \frac{8}{\pi i}\int_{\partial D_3} \biggl[k \check{\chi}_1(t,k) - \frac{12k^2 -1}{4k^2-1} \bigg(\frac{g_1}{8i} - \frac{g_0}{4\pi i}\int_{\partial \hat{D}_3} \chi_2(t,k)dk\bigg)\biggr] dk
 	\\\nonumber
& + 16\sum_{k_j \in D_1} \big(k_j - \nu_3(k_j)\big)k_j^2 e^{-2i(4k_j^3-k_j)t} \underset{k_j}{\res}R(t, k)
  	\\ \nonumber
&  - \frac{8}{\pi i} \int_{\partial D_3} e^{-2i(4k^3-k)t}  \bigl[ (\nu_1(k) - k)\nu_1^2(k)\nu_1'(k) R(t, \nu_1(k)) 
	\\\nonumber
&  + (\nu_2(k) - k)\nu_2^2(k)\nu_2'(k) R(t,\nu_2(k)) \bigr] dk
	\\  \label{N1g2expression}  
& - 2g_0^2(t) + \frac{4g_0(t)}{i\pi}\int_{\partial D_3} \hat{\chi}_2(t,k)dk+ \frac{2g_1(t)}{\pi} \int_{\partial \hat{D}_3} \chi_2(t,k)dk.
\end{align}
\end{subequations}

\item[$(c)$] For the second Neumann problem, the unknown boundary values $g_0(t)$ and $g_1(t)$ are given by
\begin{subequations}\label{N2expressions}
\begin{align}\nonumber
  g_0(t) =\;&- \frac{2}{\pi i}\int_{\partial \check{D}_3}  \frac{1}{k^2}\tilde{\chi}_1(t,k) dk
  	\\\nonumber
& - 4 \sum_{k_j \in D_1} \bigg(\frac{1}{k_j^2} - \frac{1}{\nu_3^2(k_j)}\bigg)k_j^3 e^{-2i(4k_j^3-k_j)t} \underset{k_j}{\res}R(t, k)
	\\\nonumber
&  + \frac{2}{\pi i} \int_{\partial \check{D}_3} e^{-2i(4k^3-k)t}  \bigg[ \bigg(\frac{1}{\nu_1^2(k)} - \frac{1}{k^2}\bigg)\nu_1^3(k) \nu_1'(k) R(t, \nu_1(k)) 
	\\\label{N2g0expression}
& + \bigg(\frac{1}{\nu_2^2(k)} - \frac{1}{k^2}\bigg)\nu_2^3(k)\nu_2'(k) R(t,\nu_2(k)) \bigg] dk
\end{align}
and
\begin{align}\nonumber
  g_1(t) =\;& \frac{2g_0}{\pi} \int_{\partial \hat{D}_3} \chi_2(t,k)dk 
  + \frac{4}{\pi}\int_{\partial \check{D}_3}  \frac{1}{k} \tilde{\chi}_1(t,k) dk
 	\\\nonumber
& +8i \sum_{k_j \in D_1} \bigg(\frac{1}{k_j} - \frac{1}{\nu_3(k_j)}\bigg)k_j^3 e^{-2i(4k_j^3-k_j)t} \underset{k_j}{\res}R(t, k)
	\\\nonumber
&  - \frac{4}{\pi} \int_{\partial \check{D}_3} e^{-2i(4k^3-k)t}  \bigg[ \bigg(\frac{1}{\nu_1(k)} - \frac{1}{k}\bigg)\nu_1^3(k)\nu_1'(k) R(t, \nu_1(k)) 	
	\\ \label{N2g1expression}
&  + \bigg(\frac{1}{\nu_2(k)} - \frac{1}{k}\bigg)\nu_2^3(k)\nu_2'(k) R(t,\nu_2(k)) \bigg] dk.
\end{align}
\end{subequations}
\end{itemize}
\end{theorem}
\proofbegin
$(a)$ In order to derive (\ref{g1expression}), we note that equation (\ref{g1Phi}) expresses $g_1$ in terms of $\Phi_2^{(1)}$ and $\Phi_1^{(3)}$. Furthermore, equations (\ref{Phiabcexpansions}) and Cauchy's theorem imply
\begin{align*}
  -\frac{2i\pi}{3} \Phi_2^{(1)}(t) = 2\int_{\partial \hat{D}_2} [\Phi_2(t,k) -1] dk = \int_{\partial D_4} [\Phi_2(t,k) - 1] dk
\end{align*}
and
\begin{align*}
  -\frac{2i\pi}{3} \Phi_1^{(3)}(t) = 2\int_{\partial D_2} \biggl[k^2\Phi_1(t,k) + \frac{g_0(t)}{4}\biggr] dk 
  = \int_{\partial D_4} \biggl[k^2\Phi_1(t,k) + \frac{g_0(t)}{4}\biggr] dk.
\end{align*}
Thus,
\begin{align}\nonumber
  i\pi \Phi_2^{(1)}(t) &= -\biggl(\int_{\partial \hat{D}_2} + \int_{\partial D_4}\biggr) [\Phi_2(t,k) -1] dk
  = \biggl(\int_{\partial D_1} + \int_{\partial \hat{D}_3}\biggr) [\Phi_2(t,k) -1] dk 
  	\\ \nonumber
&  =  \int_{\partial D_3} [\Phi_2(t, \nu_1(k)) -1] \nu_1'(k) dk + \int_{\partial D_3} [\Phi_2(t,\nu_2(k)) -1] \nu_2'(k) dk 
	\\ \label{ipiPhi21}
&\quad  + \int_{\partial \hat{D}_3} [\Phi_2(t,k) -1] dk 
= \int_{\partial \hat{D}_3} \chi_2(t,k)dk.
\end{align}
In order to obtain an effective construction of the Dirichlet to Neumann map for the Dirichlet problem (see Section \ref{effectivesec}), we seek an expression for $\Phi_1^{(3)}$ depending on the eigenfunction $\Phi_1(t,k)$ only via the symmetric combination $\hat{\chi}_1(t,k)$. Therefore we write
\begin{align}\nonumber
  i\pi \Phi_1^{(3)}(t) = &\; \biggl( \int_{\partial D_1} + \int_{\partial D_3}\biggr) \biggl[k^2\Phi_1(t,k) + \frac{g_0(t)}{4}\biggr] dk
  	\\ \nonumber
  =&\;  \biggl( \int_{\partial D_1'}\frac{\nu_3(k)}{\nu_1(k)} +  \int_{\partial D_1''} \frac{\nu_3(k)}{\nu_2(k)}
 +  \int_{\partial D_3} \biggr) \biggl[k^2\Phi_1(t,k) + \frac{g_0(t)}{4}\biggr] dk + I(t)
 	\\ \nonumber
 = &\; \int_{\partial D_3} \biggl[k \hat{\chi}_1(t,k) + 
 k\bigg(\frac{\nu_1'(k)}{\nu_1(k)} + \frac{\nu_2'(k)}{\nu_2(k)} +\frac{\nu_3'(k)}{\nu_3(k)}\bigg) \frac{g_0(t)}{4}\biggr] dk + I(t)
 	\\ \label{ipiPhi13}
 = &\; \int_{\partial D_3} \biggl[k \hat{\chi}_1(t,k) + \frac{12k^2 -1}{4k^2-1} \frac{g_0(t)}{4}\biggr] dk + I(t),
\end{align}
where $I(t)$ is defined by
$$I(t) =  \bigg(\int_{\partial D_1'} \bigg(1- \frac{\nu_3(k)}{\nu_1(k)}\bigg)
+ \int_{\partial D_1''} \bigg(1- \frac{{\nu_3(k)}}{\nu_2(k)}\bigg) \bigg) \biggl[k^2\Phi_1(t,k) + \frac{g_0(t)}{4}\biggr] dk$$
and we have used the identity
\begin{align}\label{nusum1}
\frac{\nu_1'(k)}{\nu_1(k)} + \frac{\nu_2'(k)}{\nu_2(k)} +\frac{\nu_3'(k)}{\nu_3(k)}
= \frac{12k^2 -1}{4k^3-k}.
\end{align}
The next step consists of using the global relation (\ref{GRc}) to compute $I(t)$:
\begin{align}\nonumber
I(t)  = &\;
\bigg(\int_{\partial D_1'} \bigg(1- \frac{\nu_3(k)}{\nu_1(k)}\bigg)
+ \int_{\partial D_1''} \bigg(1- \frac{{\nu_3(k)}}{\nu_2(k)}\bigg) \bigg) \biggl[k^2c(t,k) + \frac{g_0(t)}{4}\biggr] dk
	\\ \label{laststep}
& - \bigg(\int_{\partial D_1'} \bigg(1- \frac{\nu_3(k)}{\nu_1(k)}\bigg)
+ \int_{\partial D_1''} \bigg(1- \frac{{\nu_3(k)}}{\nu_2(k)}\bigg) \bigg)k^2 e^{-2i(4k^3-k)t} R(t,k) dk.
\end{align}
Let $\alpha = e^{2\pi i/3}$. 
Using the asymptotics (\ref{casymptotics}) of $c(t,k)$, Cauchy's theorem, and the limits
\begin{align}\label{nulimits1}
\frac{\nu_3(k)}{\nu_1(k)} = \alpha^2 + O(k^{-2}), \qquad
\frac{\nu_3(k)}{\nu_2(k)} = \alpha + O(k^{-2}), \qquad k \to \infty,
\end{align}
we find that the first term on the right-hand side of (\ref{laststep}) equals
\begin{align*}
&-\frac{\pi i}{3}(1 - \alpha^2 + 1 - \alpha)\Phi_1^{(3)}(t) + 2\pi i \sum_{k_j \in D_1'} \bigg(1 - \frac{\nu_3(k_j)}{\nu_1(k_j)}\bigg)k_j^2 e^{-2i(4k_j^3-k_j)t} \underset{k_j}{\res}R(t, k)
	\\
& + 2\pi i \sum_{k_j \in D_1''} \bigg(1 - \frac{\nu_3(k_j)}{\nu_2(k_j)}\bigg) k_j^2 e^{-2i(4k_j^3-k_j)t} \underset{k_j}{\res}R(t, k).
\end{align*}
Thus, combining the two residue sums and using the invariance of $4k^3-k$ under $k \to \nu_j(k)$, $j = 1,2,3$, in the second term on the right-hand side of (\ref{laststep}), 
\begin{align} \nonumber
 I(t) = & -i\pi \Phi_1^{(3)}(t) + 2\pi i \sum_{k_j \in D_1} \big(k_j - \nu_3(k_j)\big)k_j e^{-2i(4k_j^3-k_j)t} \underset{k_j}{\res}R(t, k)
  	\\ \nonumber
&  - \int_{\partial D_3} e^{-2i(4k^3-k)t}  \bigl[ (\nu_1(k) - \nu_3(k))\nu_1(k)\nu_1'(k) R(t, \nu_1(k)) 
	\\\label{secondtermcomputed}
&  + (\nu_2(k) - \nu_3(k))\nu_2(k)\nu_2'(k) R(t,\nu_2(k)) \bigr] dk.
\end{align}
Equations (\ref{ipiPhi13}) and (\ref{secondtermcomputed}) imply
\begin{align*}
 \Phi_1^{(3)}(t) = &\; \frac{1}{ 2\pi i}\int_{\partial D_3}  \biggl[k \hat{\chi}_1(t,k) + \frac{12k^2 -1}{4k^2-1} \frac{g_0(t)}{4}\biggr] dk
 	\\
& + \sum_{k_j \in D_1} \big(k_j - \nu_3(k_j)\big)k_j e^{-2i(4k_j^3-k_j)t} \underset{k_j}{\res}R(t, k)
	\\
&  - \frac{1}{ 2\pi i} \int_{\partial D_3} e^{-2i(4k^3-k)t}  \bigl[ (\nu_1(k) - k)\nu_1(k)\nu_1'(k) R(t, \nu_1(k)) 
	\\
&  + (\nu_2(k) - k)\nu_2(k)\nu_2'(k) R(t,\nu_2(k)) \bigr] dk.
\end{align*}
This equation together with (\ref{g1Phi}) and (\ref{ipiPhi21}) yield (\ref{g1expression}).

In order to derive (\ref{g2expression}), we note that (\ref{g2Phi}) expresses $g_2$ in terms of $\Phi_1^{(4)}$, $\Phi_2^{(2)}$, and $\Phi_2^{(1)}$. 
Furthermore, equations (\ref{Phiabcexpansions}) and Cauchy's theorem imply
\begin{align*}
  -\frac{2i\pi}{3} \Phi_1^{(4)}(t) & = 2\int_{\partial D_2} \bigg[k^3\Phi_1(t,k) + k\frac{g_0(t)}{4} - \Phi_1^{(3)}(t)\bigg] dk 
  	\\
&  = \int_{\partial D_4}  \bigg[k^3\Phi_1(t,k) + k\frac{g_0(t)}{4} - \Phi_1^{(3)}(t)\bigg] dk
\end{align*}
and
\begin{align*}
  -\frac{2i\pi}{3} \Phi_2^{(2)}(t) & = 2\int_{\partial D_2} \big[k\Phi_2(t,k) - k - \Phi_2^{(1)}(t)\big] dk 
  	\\
&  = \int_{\partial D_4} \big[k\Phi_2(t,k) - k - \Phi_2^{(1)}(t)\big] dk.
\end{align*}
Thus,
\begin{align}\nonumber
  i\pi \Phi_2^{(2)}(t) &= -\biggl(\int_{\partial D_2} + \int_{\partial D_4}\biggr) \big[k\Phi_2(t,k) - k - \Phi_2^{(1)}(t)\big]  dk
  	\\\nonumber
&  = \biggl(\int_{\partial D_1} + \int_{\partial D_3}\biggr)\big[k\Phi_2(t,k) - k - \Phi_2^{(1)}(t)\big]  dk
  	\\ \nonumber
&  = \int_{\partial D_3} \big[\nu_1(k)\Phi_2(t,\nu_1(k)) - \nu_1(k) - \Phi_2^{(1)}(t)\big] \nu_1'(k) dk 
	\\ \nonumber
&\quad  + \int_{\partial D_3} \big[\nu_2(k)\Phi_2(t,\nu_2(k)) - \nu_2(k) - \Phi_2^{(1)}(t)\big]  \nu_2'(k) dk 
	\\\label{ipiPhi22}
&\quad+ \int_{\partial D_3} \big[k\Phi_2(t,k) - k - \Phi_2^{(1)}(t)\big]  dk 
= \int_{\partial D_3} \hat{\chi}_2(t,k)dk.
\end{align}
Similarly, 
\begin{align}\nonumber
  i\pi &\Phi_1^{(4)}(t) =  \biggl( \int_{\partial D_1} + \int_{\partial D_3}\biggr)  \bigg[k^3\Phi_1(t,k) + k\frac{g_0(t)}{4} - \Phi_1^{(3)}(t)\bigg] dk
  	\\ \nonumber
 =&\;  \biggl( \int_{\partial D_1'}\frac{\nu_3^2(k)}{\nu_1^2(k)} +  \int_{\partial D_1''} \frac{\nu_3^2(k)}{\nu_2^2(k)}
 +  \int_{\partial D_3} \biggr) \bigg[k^3\Phi_1(t,k) + k\frac{g_0(t)}{4} - \Phi_1^{(3)}(t)\bigg] dk + J(t)
 	\\ \label{ipiPhi14}
 = &\; \int_{\partial D_3} \biggl[k^2 \hat{\chi}_1(t,k) + k \frac{12k^2 -1}{4k^2-1} \frac{g_0(t)}{4} + \frac{12k^2 -1}{(4k^2 -1)^2}\Phi_1^{(3)}(t)\biggr] dk + J(t),
\end{align}
where $J(t)$ is defined by
$$J(t) =  \bigg(\int_{\partial D_1'} \bigg(1- \frac{\nu_3^2(k)}{\nu_1^2(k)}\bigg)
+ \int_{\partial D_1''} \bigg(1- \frac{{\nu_3^2(k)}}{\nu_2^2(k)}\bigg) \bigg) \biggl[k^3\Phi_1(t,k) + k\frac{g_0(t)}{4} - \Phi_1^{(3)}(t)\bigg] dk$$
and we have used (\ref{nusum1}) as well as the identity
\begin{align}\label{nusum2}
\frac{\nu_1'(k)}{\nu_1^2(k)} + \frac{\nu_2'(k)}{\nu_2^2(k)} +\frac{\nu_3'(k)}{\nu_3^2(k)}
= -\frac{12k^2 -1}{k^2(4k^2 -1)^2}.
\end{align}
The contribution from the term involving $\Phi_1^{(3)}$ on the right-hand side of (\ref{ipiPhi14}) vanishes due to Cauchy's theorem.
The next step consists of using the global relation (\ref{GRc}) to compute $J(t)$:
\begin{align}\nonumber
J(t)  = &\;
\bigg(\int_{\partial D_1'} \bigg(1- \frac{\nu_3^2(k)}{\nu_1^2(k)}\bigg)
+ \int_{\partial D_1''} \bigg(1- \frac{{\nu_3^2(k)}}{\nu_2^2(k)}\bigg) \bigg) \biggl[k^3c(t,k) + k\frac{g_0(t)}{4} - \Phi_1^{(3)}\bigg] dk
	\\ \label{laststep2}
& - \bigg(\int_{\partial D_1'} \bigg(1- \frac{\nu_3^2(k)}{\nu_1^2(k)}\bigg)
+ \int_{\partial D_1''} \bigg(1- \frac{{\nu_3^2(k)}}{\nu_2^2(k)}\bigg) \bigg) k^3 e^{-2i(4k^3-k)t} R(t,k)dk.
\end{align}
Using the asymptotics (\ref{casymptotics}) of $c(t,k)$, Cauchy's theorem, and the limits (\ref{nulimits1}), we find that the first term on the right-hand side of (\ref{laststep2}) equals
\begin{align*}
&-\frac{\pi i}{3}(1 - \alpha + 1 - \alpha^2)\Phi_1^{(4)}(t) + 2\pi i \sum_{k_j \in D_1'} \bigg(1 - \frac{\nu_3^2(k_j)}{\nu_1^2(k_j)}\bigg)k_j^3 e^{-2i(4k_j^3-k_j)t} \underset{k_j}{\res}R(t, k)
	\\
& + 2\pi i \sum_{k_j \in D_1''} \bigg(1 - \frac{\nu_3^2(k_j)}{\nu_2^2(k_j)}\bigg) k_j^3 e^{-2i(4k_j^3-k_j)t} \underset{k_j}{\res}R(t, k).
\end{align*}
Thus, combining the two residue sums and using the invariance of $4k^3-k$ under $k \to \nu_j(k)$, $j = 1,2,3$, in the second term on the right-hand side of (\ref{laststep2}), 
\begin{align} \nonumber
 J(t) = & -i\pi \Phi_1^{(4)}(t) + 2\pi i \sum_{k_j \in D_1} \big(k_j^2 - \nu_3^2(k_j)\big)k_j e^{-2i(4k_j^3-k_j)t} \underset{k_j}{\res}R(t, k)
  	\\ \nonumber
&  - \int_{\partial D_3} e^{-2i(4k^3-k)t}  \bigl[ (\nu_1^2(k) - \nu_3^2(k))\nu_1(k)\nu_1'(k) R(t, \nu_1(k)) 
	\\\label{secondtermcomputed2}
&  + (\nu_2^2(k) - \nu_3^2(k))\nu_2(k)\nu_2'(k) R(t,\nu_2(k)) \bigr] dk.
\end{align}
Equations (\ref{ipiPhi14}) and (\ref{secondtermcomputed2}) imply
\begin{align}\nonumber
 \Phi_1^{(4)}(t) = &\; \frac{1}{ 2\pi i}\int_{\partial D_3} \biggl[k^2 \hat{\chi}_1(t,k) + k\frac{12k^2 -1}{4k^2-1} \frac{g_0(t)}{4} \biggr] dk
 	\\\nonumber
& +  \sum_{k_j \in D_1} \big(k_j^2 - \nu_3^2(k_j)\big)k_j e^{-2i(4k_j^3-k_j)t} \underset{k_j}{\res}R(t, k)
  	\\ \nonumber
&  - \frac{1}{ 2\pi i} \int_{\partial D_3} e^{-2i(4k^3-k)t}  \bigl[ (\nu_1^2(k) - k^2)\nu_1(k)\nu_1'(k) R(t, \nu_1(k)) 
	\\\label{Phi14}
& + (\nu_2^2(k) - k^2)\nu_2(k)\nu_2'(k) R(t,\nu_2(k)) \bigr] dk.	
\end{align}
Substituting the expressions (\ref{ipiPhi21}), (\ref{ipiPhi22}), and (\ref{Phi14}) for $\Phi_2^{(1)}$, $\Phi_2^{(2)}$, $\Phi_1^{(4)}$ into (\ref{g2Phi}) we obtain (\ref{g2expression}).

$(b)$ In order to derive the representation (\ref{N1g0expression}) for $g_0$ relevant for the first Neumann problem, we note that equation (\ref{g0Phi}) expresses $g_0$ in terms of $\Phi_1^{(2)}$. Furthermore, equations (\ref{Phiabcexpansions}) and Cauchy's theorem imply
\begin{align*}
  -\frac{2i\pi}{3} \Phi_1^{(2)}(t) = 2\int_{\partial D_2} k\Phi_1(t,k) dk = \int_{\partial D_4} k\Phi_1(t,k) dk.
\end{align*}
Thus,
\begin{align}\nonumber
  i\pi \Phi_1^{(2)}(t) =& -\biggl(\int_{\partial D_2} + \int_{\partial D_4}\biggr) k\Phi_1(t,k) dk
  = \biggl(\int_{\partial D_1} + \int_{\partial D_3}\biggr) k\Phi_1(t,k) dk
  	\\ \nonumber
 =&\;  \biggl( \int_{\partial \check{D}_1'}\frac{\nu_1(k)}{\nu_3(k)} +  \int_{\partial \check{D}_1''} \frac{\nu_2(k)}{\nu_3(k)}
 +  \int_{\partial D_3} \biggr) k\Phi_1(t,k) dk + I(t)
 	\\  \label{N1ipiPhi12}
 = &\; \int_{\partial \check{D}_3} \frac{1}{k}\check{\chi}_1(t,k)dk + I(t),
\end{align}
where $I(t)$ is defined by
$$I(t) =  \bigg(\int_{\partial \check{D}_1'} \bigg(1- \frac{\nu_1(k)}{\nu_3(k)}\bigg)
+ \int_{\partial \check{D}_1''} \bigg(1- \frac{{\nu_2(k)}}{\nu_3(k)}\bigg) \bigg) k\Phi_1(t,k) dk.$$
The next step consists of using the global relation (\ref{GRc}) to compute $I(t)$:
\begin{align}\nonumber
I(t)  = &\;
\bigg(\int_{\partial \check{D}_1'} \bigg(1- \frac{\nu_1(k)}{\nu_3(k)}\bigg)
+ \int_{\partial \check{D}_1''} \bigg(1- \frac{{\nu_2(k)}}{\nu_3(k)}\bigg) \bigg) kc(t,k) dk
	\\ \label{N1laststep}
& - \bigg(\int_{\partial \check{D}_1'} \bigg(1- \frac{\nu_1(k)}{\nu_3(k)}\bigg)
+ \int_{\partial \check{D}_1''} \bigg(1- \frac{{\nu_2(k)}}{\nu_3(k)}\bigg) \bigg)k e^{-2i(4k^3-k)t} R(t,k) dk.
\end{align}
Using the asymptotics (\ref{casymptotics}) of $c(t,k)$, Cauchy's theorem, and the limits (\ref{nulimits1}), we find that the first term on the right-hand side of (\ref{N1laststep}) equals
\begin{align*}
&-\frac{\pi i}{3}(1 - \alpha + 1 - \alpha^2)\Phi_1^{(2)} + 2\pi i \sum_{k_j \in D_1'} \bigg(1 - \frac{\nu_1(k_j)}{\nu_3(k_j)}\bigg)k_j e^{-2i(4k_j^3-k_j)t} \underset{k_j}{\res}R(t, k)
	\\
& + 2\pi i \sum_{k_j \in D_1''} \bigg(1 - \frac{\nu_2(k_j)}{\nu_3(k_j)}\bigg) k_j e^{-2i(4k_j^3-k_j)t} \underset{k_j}{\res}R(t, k).
\end{align*}
Thus, 
\begin{align} \nonumber
 I(t) = & -i\pi \Phi_1^{(2)}(t) + 2\pi i \sum_{k_j \in D_1} \bigg(\frac{1}{k_j} - \frac{1}{\nu_3(k_j)}\bigg)k_j^2 e^{-2i(4k_j^3-k_j)t} \underset{k_j}{\res}R(t, k)
  	\\ \nonumber
&  - \int_{\partial \check{D}_3} e^{-2i(4k^3-k)t}  \bigg[ \bigg(\frac{1}{\nu_1(k)} - \frac{1}{\nu_3(k)}\bigg)\nu_1^2(k) \nu_1'(k) R(t, \nu_1(k)) 
	\\\label{N1secondtermcomputed}
& + \bigg(\frac{1}{\nu_2(k)} - \frac{1}{\nu_3(k)}\bigg)\nu_2^2(k)\nu_2'(k) R(t,\nu_2(k)) \bigg] dk.
\end{align}
Equations (\ref{N1ipiPhi12}) and (\ref{N1secondtermcomputed}) imply
\begin{align*}
 \Phi_1^{(2)}(t) = &\; \frac{1}{ 2\pi i}\int_{\partial \check{D}_3}  \frac{1}{k}\check{\chi}_1(t,k) dk
 +  \sum_{k_j \in D_1} \bigg(\frac{1}{k_j} - \frac{1}{\nu_3(k_j)}\bigg)k_j^2 e^{-2i(4k_j^3-k_j)t} \underset{k_j}{\res}R(t, k)
	\\
&  - \frac{1}{ 2\pi i} \int_{\partial \check{D}_3} e^{-2i(4k^3-k)t}  \bigg[ \bigg(\frac{1}{\nu_1(k)} - \frac{1}{k}\bigg)\nu_1^2(k) \nu_1'(k) R(t, \nu_1(k)) 
	\\
& + \bigg(\frac{1}{\nu_2(k)} - \frac{1}{k}\bigg)\nu_2^2(k)\nu_2'(k) R(t,\nu_2(k)) \bigg] dk.
\end{align*}
In view of (\ref{g0Phi}), this yields (\ref{N1g0expression}).

In order to derive (\ref{N1g2expression}), we note that (\ref{ipiPhi14}) implies
\begin{align}\nonumber
&  i\pi \Phi_1^{(4)}(t) =  \biggl( \int_{\partial D_1} + \int_{\partial D_3}\biggr)  \bigg[k^3\Phi_1(t,k) + k\frac{g_0(t)}{4} - \Phi_1^{(3)}(t)\bigg] dk
  	\\ \nonumber
 =&\;  \biggl( \int_{\partial D_1'}\frac{\nu_3(k)}{\nu_1(k)} +  \int_{\partial D_1''} \frac{\nu_3(k)}{\nu_2(k)}
 +  \int_{\partial D_3} \biggr) \bigg[k^3\Phi_1(t,k) + k\frac{g_0(t)}{4} - \Phi_1^{(3)}(t)\bigg] dk + J(t)
 	\\ \label{N1ipiPhi14}
 = &\; \int_{\partial D_3} \biggl[k \check{\chi}_1(t,k) - k \frac{12k^2 -1}{4k^3-k} \Phi_1^{(3)}(t)\biggr] dk + J(t),
\end{align}
where $J(t)$ is defined by
$$J(t) =  \bigg(\int_{\partial D_1'} \bigg(1- \frac{\nu_3(k)}{\nu_1(k)}\bigg)
+ \int_{\partial D_1''} \bigg(1- \frac{{\nu_3(k)}}{\nu_2(k)}\bigg) \bigg) \biggl[k^3\Phi_1(t,k) + k\frac{g_0(t)}{4} - \Phi_1^{(3)}(t)\bigg] dk$$
and we have used the identity (\ref{nusum1}).
The next step consists of using the global relation (\ref{GRc}) to compute $J(t)$:
\begin{align}\nonumber
J(t)  = &\;
\bigg(\int_{\partial D_1'} \bigg(1- \frac{\nu_3(k)}{\nu_1(k)}\bigg)
+ \int_{\partial D_1''} \bigg(1- \frac{{\nu_3(k)}}{\nu_2(k)}\bigg) \bigg) \biggl[k^3c(t,k) + k\frac{g_0(t)}{4} - \Phi_1^{(3)}(t)\bigg] dk
	\\ \label{N1laststep2}
& - \bigg(\int_{\partial D_1'} \bigg(1- \frac{\nu_3(k)}{\nu_1(k)}\bigg)
+ \int_{\partial D_1''} \bigg(1- \frac{{\nu_3(k)}}{\nu_2(k)}\bigg) \bigg) k^3 e^{-2i(4k^3-k)t} R(t,k)dk.
\end{align}
Using the asymptotics (\ref{casymptotics}) of $c(t,k)$, Cauchy's theorem, and the limits (\ref{nulimits1}), we find that the first term on the right-hand side of (\ref{N1laststep2}) equals
\begin{align*}
&-\frac{\pi i}{3}(1 - \alpha^2 + 1 - \alpha)\Phi_1^{(4)}(t) + 2\pi i \sum_{k_j \in D_1'} \bigg(1 - \frac{\nu_3(k_j)}{\nu_1(k_j)}\bigg)k_j^3 e^{-2i(4k_j^3-k_j)t} \underset{k_j}{\res}R(t, k)
	\\
& + 2\pi i \sum_{k_j \in D_1''} \bigg(1 - \frac{\nu_3(k_j)}{\nu_2(k_j)}\bigg) k_j^3 e^{-2i(4k_j^3-k_j)t} \underset{k_j}{\res}R(t, k).
\end{align*}
Thus, 
\begin{align} \nonumber
 J(t) = &\; -i\pi \Phi_1^{(4)}(t) + 2\pi i \sum_{k_j \in D_1} \big(k_j - \nu_3(k_j)\big)k_j^2 e^{-2i(4k_j^3-k_j)t} \underset{k_j}{\res}R(t, k)
  	\\ \nonumber
&  - \int_{\partial D_3} e^{-2i(4k^3-k)t}  \bigl[ (\nu_1(k) - \nu_3(k))\nu_1^2(k)\nu_1'(k) R(t, \nu_1(k)) 
	\\\label{N1secondtermcomputed2}
& + (\nu_2(k) - \nu_3(k))\nu_2^2(k)\nu_2'(k) R(t,\nu_2(k)) \bigr] dk.
\end{align}
Equations (\ref{N1ipiPhi14}) and (\ref{N1secondtermcomputed2}) imply
\begin{align}\nonumber
 \Phi_1^{(4)}(t) = &\; \frac{1}{ 2\pi i}\int_{\partial D_3} \biggl[k \check{\chi}_1(t,k) - k \frac{12k^2 -1}{4k^3-k} \bigg(\frac{g_1}{8i} - \frac{g_0}{4}\Phi_2^{(1)}(t)\bigg)\biggr] dk
 	\\\nonumber
& + \sum_{k_j \in D_1} \big(k_j - \nu_3(k_j)\big)k_j^2 e^{-2i(4k_j^3-k_j)t} \underset{k_j}{\res}R(t, k)
  	\\ \nonumber
&  - \frac{1}{ 2\pi i} \int_{\partial D_3} e^{-2i(4k^3-k)t}  \bigl[ (\nu_1(k) - k)\nu_1^2(k)\nu_1'(k) R(t, \nu_1(k)) 
	\\\label{N1Phi14}
& + (\nu_2(k) - k)\nu_2^2(k)\nu_2'(k) R(t,\nu_2(k)) \bigr] dk.	
\end{align}
Substituting the expressions (\ref{ipiPhi21}), (\ref{ipiPhi22}), and (\ref{N1Phi14}) for $\Phi_2^{(1)}$, $\Phi_2^{(2)}$, $\Phi_1^{(4)}$ into (\ref{g2Phi}) we obtain (\ref{N1g2expression}).

$(c)$ In order to derive the representation (\ref{N2g0expression}) for $g_0$ relevant for the second Neumann problem, we note that (\ref{N1ipiPhi12}) implies
\begin{align}\nonumber
  i\pi \Phi_1^{(2)}(t) =& \biggl(\int_{\partial D_1} + \int_{\partial D_3}\biggr) k\Phi_1(t,k) dk
  	\\ \nonumber
 =&\;  \biggl( \int_{\partial \check{D}_1'}\frac{\nu_1^2(k)}{\nu_3^2(k)} 
 + \int_{\partial \check{D}_1''} \frac{\nu_2^2(k)}{\nu_3^2(k)}
 + \int_{\partial D_3} \biggr) k\Phi_1(t,k) dk + I(t)
 	\\  \label{N2ipiPhi12}
 = &\; \int_{\partial D_3} \frac{1}{k^2}\tilde{\chi}_1(t,k)dk + I(t),
\end{align}
where $I(t)$ is defined by
$$I(t) =  \bigg(\int_{\partial \check{D}_1'} \bigg(1- \frac{\nu_1^2(k)}{\nu_3^2(k)}\bigg)
+ \int_{\partial \check{D}_1''} \bigg(1- \frac{{\nu_2^2(k)}}{\nu_3^2(k)}\bigg) \bigg) k\Phi_1(t,k) dk.$$
The next step consists of using the global relation (\ref{GRc}) to compute $I(t)$:
\begin{align}\nonumber
I(t)  = &\;
\bigg(\int_{\partial \check{D}_1'} \bigg(1- \frac{\nu_1^2(k)}{\nu_3^2(k)}\bigg)
+ \int_{\partial \check{D}_1''} \bigg(1- \frac{\nu_2^2(k)}{\nu_3^2(k)}\bigg) \bigg) kc(t,k) dk
	\\ \label{N2laststep}
& - \bigg(\int_{\partial \check{D}_1'} \bigg(1- \frac{\nu_1^2(k)}{\nu_3^2(k)}\bigg)
+ \int_{\partial \check{D}_1''} \bigg(1- \frac{{\nu_2^2(k)}}{\nu_3^2(k)}\bigg) \bigg)k e^{-2i(4k^3-k)t} R(t,k) dk.
\end{align}
Using the asymptotics (\ref{casymptotics}) of $c(t,k)$, Cauchy's theorem, and the limits (\ref{nulimits1}), we find that the first term on the right-hand side of (\ref{N2laststep}) equals
\begin{align*}
&-\frac{\pi i}{3}(1 - \alpha^2 + 1 - \alpha)\Phi_1^{(2)}(t) + 2\pi i \sum_{k_j \in D_1'} \bigg(1 - \frac{\nu_1^2(k_j)}{\nu_3^2(k_j)}\bigg)k_j e^{-2i(4k_j^3-k_j)t} \underset{k_j}{\res}R(t, k)
	\\
& + 2\pi i \sum_{k_j \in D_1''} \bigg(1 - \frac{\nu_2^2(k_j)}{\nu_3^2(k_j)}\bigg) k_j e^{-2i(4k_j^3-k_j)t} \underset{k_j}{\res}R(t, k).
\end{align*}
Thus, 
\begin{align} \nonumber
 I(t) = & -i\pi \Phi_1^{(2)}(t) + 2\pi i \sum_{k_j \in D_1} \bigg(\frac{1}{k_j^2} - \frac{1}{\nu_3^2(k_j)}\bigg)k_j^3 e^{-2i(4k_j^3-k_j)t} \underset{k_j}{\res}R(t, k)
  	\\ \nonumber
&  - \int_{\partial \check{D}_3} e^{-2i(4k^3-k)t}  \bigg[ \bigg(\frac{1}{\nu_1^2(k)} - \frac{1}{\nu_3^2(k)}\bigg)\nu_1^3(k) \nu_1'(k) R(t, \nu_1(k)) 
	\\\label{N2secondtermcomputed}
& + \bigg(\frac{1}{\nu_2^2(k)} - \frac{1}{\nu_3^2(k)}\bigg)\nu_2^3(k)\nu_2'(k) R(t,\nu_2(k)) \bigg] dk.
\end{align}
Equations (\ref{N2ipiPhi12}) and (\ref{N2secondtermcomputed}) imply
\begin{align*}
 \Phi_1^{(2)}(t) = &\; \frac{1}{ 2\pi i}\int_{\partial \check{D}_3}  \frac{1}{k^2}\tilde{\chi}_1(t,k) dk
 +  \sum_{k_j \in D_1} \bigg(\frac{1}{k_j^2} - \frac{1}{\nu_3^2(k_j)}\bigg)k_j^3 e^{-2i(4k_j^3-k_j)t} \underset{k_j}{\res}R(t, k)
	\\
&  - \frac{1}{ 2\pi i} \int_{\partial \check{D}_3} e^{-2i(4k^3-k)t}  \bigg[ \bigg(\frac{1}{\nu_1^2(k)} - \frac{1}{k^2}\bigg)\nu_1^3(k) \nu_1'(k) R(t, \nu_1(k)) 
	\\
& + \bigg(\frac{1}{\nu_2^2(k)} - \frac{1}{k^2}\bigg)\nu_2^3(k)\nu_2'(k) R(t,\nu_2(k)) \bigg] dk.
\end{align*}
In view of (\ref{g0Phi}), this yields (\ref{N2g0expression}).

 In order to derive (\ref{N2g1expression}), we note that (\ref{ipiPhi13}) implies
 \begin{align}\nonumber
  i\pi \Phi_1^{(3)}(t) = &\; \biggl( \int_{\partial D_1} + \int_{\partial D_3}\biggr) \biggl[k^2\Phi_1(t,k) + \frac{g_0(t)}{4}\biggr] dk
  	\\ \nonumber
 =&\;  \biggl( \int_{\partial \check{D}_1'}\frac{\nu_1(k)}{\nu_3(k)} +  \int_{\partial \check{D}_1''} \frac{\nu_2(k)}{\nu_3(k)}
 +  \int_{\partial D_3} \biggr) \biggl[k^2\Phi_1(t,k) + \frac{g_0(t)}{4}\biggr] dk + J(t)
 	\\ \label{N2ipiPhi13}
 = &\; \int_{\partial \check{D}_3} \frac{1}{k}\tilde{\chi}_1(t,k) dk + J(t),
\end{align}
where $J(t)$ is defined by
$$J(t) =  \bigg(\int_{\partial \check{D}_1'} \bigg(1- \frac{\nu_1(k)}{\nu_3(k)}\bigg)
+ \int_{\partial \check{D}_1''} \bigg(1- \frac{{\nu_2(k)}}{\nu_3(k)}\bigg) \bigg) \biggl[k^2\Phi_1(t,k) + \frac{g_0(t)}{4}\biggr] dk.$$
The next step consists of using the global relation (\ref{GRc}) to compute $J(t)$:
\begin{align}\nonumber
J(t)  = &\;
\bigg(\int_{\partial \check{D}_1'} \bigg(1- \frac{\nu_1(k)}{\nu_3(k)}\bigg)
+ \int_{\partial \check{D}_1''} \bigg(1- \frac{{\nu_2(k)}}{\nu_3(k)}\bigg) \bigg) \biggl[k^2c(t,k) + \frac{g_0(t)}{4}\biggr] dk
	\\ \label{N2laststep2}
& - \bigg(\int_{\partial \check{D}_1'} \bigg(1- \frac{\nu_1(k)}{\nu_3(k)}\bigg)
+ \int_{\partial \check{D}_1''} \bigg(1- \frac{{\nu_2(k)}}{\nu_3(k)}\bigg) \bigg)k^2 e^{-2i(4k^3-k)t} R(t,k) dk.
\end{align}
Using the asymptotics (\ref{casymptotics}) of $c(t,k)$, Cauchy's theorem, and the limits (\ref{nulimits1}), we find that the first term on the right-hand side of (\ref{N2laststep2}) equals
\begin{align*}
&-\frac{\pi i}{3}(1 - \alpha + 1 - \alpha^2)\Phi_1^{(3)}(t) + 2\pi i \sum_{k_j \in D_1'} \bigg(1 - \frac{\nu_1(k_j)}{\nu_3(k_j)}\bigg)k_j^2 e^{-2i(4k_j^3-k_j)t} \underset{k_j}{\res}R(t, k)
	\\
& + 2\pi i \sum_{k_j \in D_1''} \bigg(1 - \frac{\nu_2(k_j)}{\nu_3(k_j)}\bigg) k_j^2 e^{-2i(4k_j^3-k_j)t} \underset{k_j}{\res}R(t, k).
\end{align*}
Thus, 
\begin{align} \nonumber
 J(t) = &\; -i\pi \Phi_1^{(3)}(t) + 2\pi i \sum_{k_j \in D_1} \bigg(\frac{1}{k_j} - \frac{1}{\nu_3(k_j)}\bigg)k_j^3 e^{-2i(4k_j^3-k_j)t} \underset{k_j}{\res}R(t, k)
  	\\ \nonumber
&  - \int_{\partial \check{D}_3} e^{-2i(4k^3-k)t}  \bigg[ \bigg(\frac{1}{\nu_1(k)} - \frac{1}{\nu_3(k)}\bigg)\nu_1^3(k)\nu_1'(k) R(t, \nu_1(k)) 
	\\\label{N2secondtermcomputed2}
&  + \bigg(\frac{1}{\nu_2(k)} - \frac{1}{\nu_3(k)}\bigg)\nu_2^3(k)\nu_2'(k) R(t,\nu_2(k)) \bigg] dk.
\end{align}
Equations (\ref{N2ipiPhi13}) and (\ref{N2secondtermcomputed2}) imply
\begin{align*}
 \Phi_1^{(3)}(t) = &\; \frac{1}{ 2\pi i}\int_{\partial \check{D}_3}  \frac{1}{k} \tilde{\chi}_1(t,k) dk
 	\\
& + \sum_{k_j \in D_1} \bigg(\frac{1}{k_j} - \frac{1}{\nu_3(k_j)}\bigg)k_j^3 e^{-2i(4k_j^3-k_j)t} \underset{k_j}{\res}R(t, k)
	\\
&  - \frac{1}{ 2\pi i} \int_{\partial \check{D}_3} e^{-2i(4k^3-k)t}  \bigg[ \bigg(\frac{1}{\nu_1(k)} - \frac{1}{k}\bigg)\nu_1^3(k)\nu_1'(k) R(t, \nu_1(k)) 	
	\\
&  + \bigg(\frac{1}{\nu_2(k)} - \frac{1}{k}\bigg)\nu_2^3(k)\nu_2'(k) R(t,\nu_2(k)) \bigg] dk.
\end{align*}
This equation together with (\ref{g1Phi}) and (\ref{ipiPhi21}) yield (\ref{N2g1expression}).
\proofend

\section{Effective characterization}\nequation\label{effectivesec}
The nonlinear system for $\Phi_1$ andÊ $\Phi_2$ obtained by substituting the expressions (\ref{g1g2expression}) for $g_1$ and $g_2$ into (\ref{Phieqs}) provides an effective characterization of the Dirichlet to Neumann map for the Dirichlet problem for the KdV equation. 
Similarly, substituting the representations (\ref{N1expressions}) and (\ref{N2expressions}) into (\ref{Phieqs}) yields an effective characterization of the (generalized) Dirichlet to Neumann map for the first and second Neumann problems respectively.
Indeed, substituting into the system (\ref{Phieqs}) the expansions
\begin{align*}
 \Phi_j &= \Phi_{j0} + \epsilon \Phi_{j1} + \epsilon^2 \Phi_{j2} + \cdots, \qquad j = 1,2,
  	\\
 g_j &= \epsilon g_{j1} + \epsilon^2 g_{j2} + \cdots, \qquad j = 0, 1,2,
\end{align*}
where $\epsilon > 0$ is a small parameter, we find that the terms of $O(1)$ give $\Phi_{10} \equiv 0$ and $\Phi_{20} \equiv 1$, while the terms of $O(\epsilon)$ give
\begin{align}\nonumber
& \Phi_{11}(t, k) = \int_0^t e^{2i(4k^3-k)(t' - t)}\left(-2ikg_{01}(t') + g_{11}(t') + \frac{i}{2k}(g_{01}(t') + g_{21}(t'))\right)dt',
	\\ \label{eps1a}
& \Phi_{21}(t,k) = - \frac{i}{2k} \int_0^t (g_{01}(t') + g_{21}(t'))dt',
\end{align}
We let $\chi_j = \chi_{j1} \epsilon + \chi_{j2} \epsilon^2 + \cdots$ denote the perturbative expansion of $\chi_j$, $j = 1,2$, and adopt similar notation for the other symmetric combinations defined in (\ref{chidef}). Then, using the following identities which are a consequence of (\ref{nurelations}):
\begin{align*}
&\sum_{j=1}^3 \frac{1}{\nu_j(k)} = -\frac{1}{4k^3 - k}, \qquad
\sum_{j=1}^3 \frac{\nu_j'(k)}{\nu_j(k)} = \frac{12k^2 -1}{4k^3-k},
	\\
&\sum_{j=1}^3 \nu_j^2(k)\nu_j'(k) = \frac{12k^2 -1}{4},
\qquad \sum_{j=1}^3 \nu_j^3(k)\nu_j'(k) = 0,
\qquad \sum_{j=1}^3 \nu_j^4(k)\nu_j'(k) = \frac{12k^2 -1}{16},
\end{align*}
we discover that
\begin{subequations}
\begin{align}\label{chi11}
& \chi_{11}(t, k) = \frac{i}{2} \frac{12k^2 -1}{4k^3-k} \int_0^t e^{2i(4k^3-k)(t' - t)}(g_{01}(t') + g_{21}(t'))dt',
	\\ \label{hatchi11}
& \hat{\chi}_{11}(t, k) = -\frac{i}{2} (12k^2 -1) \int_0^t e^{2i(4k^3-k)(t' - t)} g_{01}(t') dt',
	\\ \label{checkchi11}
& \check{\chi}_{11}(t, k) = \frac{12 k^2 - 1}{4} \int_0^t e^{2i(4k^3-k)(t' - t)} g_{11}(t') dt',
	\\ \label{tildechi11}
& \tilde{\chi}_{11}(t,k) = i\frac{12k^2 -1}{8} \int_0^t e^{2i(4k^3-k)(t' - t)}  g_{21}(t') dt'.	
\end{align}
\end{subequations}

\subsection{The Dirichlet problem}
The Dirichlet problem can now be solved perturbatively as follows. 
Let
$$a = 1 + a_1 \epsilon + a_2 \epsilon^2 + \cdots, \qquad
b = b_1 \epsilon + b_2\epsilon^2 + \cdots, \qquad \epsilon \to 0,$$
denote the expansions of $a(k)$Ê and $b(k)$.
Expanding (\ref{g1expression}) and (\ref{g2expression}) and assuming that $a(k)$ has no zeros (note that $a(k) \simeq 1$ in the linear limit), we find
\begin{subequations}\label{g11g21expressions}
\begin{align} \nonumber
g_{11}(t) = \;& \frac{4}{\pi}\int_{\partial D_3}  \biggl[k \hat{\chi}_{11}(t,k) + \frac{12k^2 -1}{4k^2-1} \frac{g_{01}(t)}{4}\biggr] dk
 	\\\nonumber
&  - \frac{4}{ \pi } \int_{\partial D_3} e^{-2i(4k^3-k)t}  \bigl[ (\nu_1(k) - k)\nu_1(k)\nu_1'(k) b_1(\nu_1(k)) 
	\\\label{g11expression}
&  + (\nu_2(k) - k)\nu_2(k)\nu_2'(k) b_1(\nu_2(k)) \bigr] dk,
	\\ \nonumber
g_{21}(t) = \;& 
\frac{8}{\pi i}\int_{\partial D_3} \biggl[k^2 \hat{\chi}_{11}(t,k) + k\frac{12k^2 -1}{4k^2-1} \frac{g_{01}(t)}{4} \biggr] dk
  	\\ \nonumber
& - \frac{8}{\pi i} \int_{\partial D_3} e^{-2i(4k^3-k)t}  \bigl[ (\nu_1^2(k) - k^2)\nu_1(k)\nu_1'(k) b_1(\nu_1(k)) 
	\\ \label{g21expression}
&  + (\nu_2^2(k) - k^2)\nu_2(k)\nu_2'(k) b_1(\nu_2(k)) \bigr] dk.
\end{align}
\end{subequations}
Using equation (\ref{hatchi11}) to determine $\hat{\chi}_{11}$, we can determine $g_{11}$, $g_{21}$ from (\ref{g11g21expressions}); then $\Phi_{11}$ and $\Phi_{21}$ can be found from (\ref{eps1a}). 
These arguments can be extended to higher orders and thus yield a constructive scheme for computing the Dirichlet to Neumann map to all orders. Indeed, the terms of order $O(\epsilon^n)$ yield
\begin{subequations}
\begin{align}
& \chi_{1n}(t, k) = \frac{i}{2} \frac{12k^2 -1}{4k^3-k} \int_0^t e^{2i(4k^3-k)(t' - t)}(g_{0n}(t') + g_{2n}(t'))dt' + \lot,
	\\ 
& \hat{\chi}_{1n}(t, k) = -\frac{i}{2} (12k^2 -1) \int_0^t e^{2i(4k^3-k)(t' - t)} g_{0n}(t') dt' + \lot,
	\\ \label{checkchi1n}
& \check{\chi}_{1n}(t, k) = \frac{12 k^2 - 1}{4} \int_0^t e^{2i(4k^3-k)(t' - t)} g_{1n}(t') dt' + \lot,
	\\ \label{tildechi1n}
& \tilde{\chi}_{1n}(t,k) = i\frac{12k^2 -1}{8} \int_0^t e^{2i(4k^3-k)(t' - t)}  g_{2n}(t') dt' + \lot,
\end{align}
\end{subequations}
where `$\lot$' denotes an expression involving known terms of lower order. Moreover, (\ref{g1g2expression}) yields
\begin{align*} \nonumber
g_{1n}(t) = \;& \frac{4}{\pi}\int_{\partial D_3}  \biggl[k \hat{\chi}_{1n}(t,k) + \frac{12k^2 -1}{4k^2-1} \frac{g_{0n}(t)}{4}\biggr] dk
 	\\\nonumber
&  - \frac{4}{ \pi } \int_{\partial D_3} e^{-2i(4k^3-k)t}  \bigl[ (\nu_1(k) - k)\nu_1(k)\nu_1'(k) b_n(\nu_1(k)) 
	\\
& + (\nu_2(k) - k)\nu_2(k)\nu_2'(k) b_n(\nu_2(k)) \bigr] dk + \lot,
	\\ \nonumber
g_{2n}(t) = \;& 
\frac{8}{\pi i}\int_{\partial D_3} \biggl[k^2 \hat{\chi}_{1n}(t,k) + k\frac{12k^2 -1}{4k^2-1} \frac{g_{0n}(t)}{4} \biggr] dk
  	\\ \nonumber
& - \frac{8}{\pi i} \int_{\partial D_3} e^{-2i(4k^3-k)t}  \bigl[ (\nu_1^2(k) - k^2)\nu_1(k)\nu_1'(k) b_n(\nu_1(k)) 
	\\
& + (\nu_2^2(k) - k^2)\nu_2(k)\nu_2'(k) b_n(\nu_2(k)) \bigr] dk + \lot.
\end{align*}
Since $\hat{\chi}_{1n}$ and $b_n$ can be computed from the given initial and Dirichlet boundary conditions, we can find $g_{1n}$ and $g_{2n}$ and then proceed to the next order.

\subsection{The first Neumann problem}
For the first Neumann problem, the terms of $O(\epsilon^n)$ of (\ref{N1expressions}) yield
\begin{subequations}\label{N1pertexpression}
\begin{align}\nonumber
  g_{0n}(t) =& -\frac{2}{\pi i}\int_{\partial \check{D}_3}  \frac{1}{k}\check{\chi}_{1n}(t,k) dk
 	\\\nonumber
&+ \frac{2}{\pi i} \int_{\partial \check{D}_3} e^{-2i(4k^3-k)t}  \bigg[ \bigg(\frac{1}{\nu_1(k)} - \frac{1}{k}\bigg)\nu_1^2(k) \nu_1'(k) b_n(\nu_1(k)) 
	\\   
& + \bigg(\frac{1}{\nu_2(k)} - \frac{1}{k}\bigg)\nu_2^2(k)\nu_2'(k) b_n(\nu_2(k)) \bigg] dk + \lot,
	\\ \nonumber
  g_{2n}(t) =\;&
 \frac{8}{\pi i}\int_{\partial D_3} \biggl[k \check{\chi}_{1n}(t,k) - \frac{12k^2 -1}{4k^2-1} \frac{g_{1n}}{8i} \biggr] dk
  	\\ \nonumber
&  - \frac{8}{\pi i} \int_{\partial D_3} e^{-2i(4k^3-k)t}  \bigl[ (\nu_1(k) - k)\nu_1^2(k)\nu_1'(k) b_n(\nu_1(k)) 
	\\
& + (\nu_2(k) - k)\nu_2^2(k)\nu_2'(k) b_n(\nu_2(k)) \bigr] dk + \lot.
\end{align}
\end{subequations}
In this case, we use equation (\ref{checkchi1n}) to determine $\check{\chi}_{1n}$; we then determine $g_{0n}$, $g_{2n}$ from (\ref{N1pertexpression}); then $\Phi_{1n}$ and $\Phi_{2n}$ can be found from the $O(\epsilon^n)$ terms of the system (\ref{Phieqs}).

\subsection{The second Neumann problem}
For the second Neumann problem, a similar argument shows that the terms of $O(\epsilon^n)$ of (\ref{N2expressions}) yield expressions for $g_{0n}$ and $g_{1n}$ in terms of $\tilde{\chi}_{1n}$ and $b_n$ as well as lower order terms. Since $\tilde{\chi}_{1n}$ can be computed from (\ref{tildechi1n}) in terms of the given Neumann data $g_2$ alone, we can proceed to the next order.

\bigskip
\noindent
{\bf Acknowledgement} {\it The author is grateful for support from the EPSRC, UK.}

\bibliographystyle{plain}
\bibliography{is}

\begin{thebibliography}{99}
\small

\bibitem{BPS1981}
J. L. Bona, W. G. Pritchard, and L. R. Scott,
An evaluation of a model equation for water waves, {\it Philos. Trans. Roy. Soc. London Ser. A} {\bf 302} (1981), 457--510. 

\bibitem{BSZ2002}
J. Bona, S. M. Sun, and B.-Y. Zhang, A non-homogeneous boundary-value problem for the Korteweg-de Vries equation in a quarter plane, {\it Trans. Amer. Math. Soc.} {\bf 354} (2002), 427--490.

\bibitem{BSZ2008}
J. Bona, S. M. Sun, and B.-Y. Zhang, Non-homogeneous boundary value problems for the Korteweg-de Vries and the Korteweg-de Vries-Burgers equations in a quarter plane, {\it Ann. Inst. H. PoincarŽ Anal. Non Lin\'eaire} {\bf 25} (2008), 1145--1185.

\bibitem{BW1983}
J. Bona and R. Winther, The Korteweg-de Vries equation, posed in a quarter-plane, {\it 
SIAM J. Math. Anal.} {\bf 14} (1983), 1056--1106. 

\bibitem{BFS2003}
A. Boutet de Monvel, A. S. Fokas, and D. Shepelsky, The analysis of the
global relation for the nonlinear Schr\"odinger equation on the half-line,
{\it Lett. Math. Phys.} {\bf 65} (2003), 199--212.

\bibitem{DT1979}
P. Deift and E. Trubowitz, Inverse scattering on the line, {\it Comm. Pure Appl. Math.} {\bf 32} (1979), 121--251.

\bibitem{F1997}
A. S. Fokas, A unified transform method for solving linear and certain nonlinear PDEs, 
{\it Proc. Roy. Soc. Lond.} A {\bf 453} (1997), 1411--1443.

\bibitem{F2002}
A. S. Fokas, Integrable nonlinear evolution equations on the half-line, 
{\it Comm. Math. Phys.} {\bf 230} (2002), 1--39.

\bibitem{F2005}
A. S. Fokas, A generalised Dirichlet to Neumann map for certain nonlinear evolution PDEs, 
{\it Comm. Pure Appl. Math.} {\bf LVIII} (2005), 639--670.

\bibitem{trilogy1}
A. S. Fokas and J. Lenells, The unified method: I Non-linearizable problems on the half-line, {\it J. Phys. A: Math. Theor.} {\bf 45} (2012), 195201.

\bibitem{LdnlsD2N}
J. Lenells, The solution of the global relation for the derivative nonlinear Schr\"odinger equation on the half-line, {\it Physica D} {\bf 240} (2011), 512--525.

\bibitem{L3x3}
J. Lenells, Initial-boundary value problems for integrable evolution equations with $3\times 3$ Lax pairs, {\it Physica D} {\bf 241} (2012), 857--875.

\bibitem{trilogy3}
J. Lenells and A. S. Fokas, The unified method: III Non-linearizable problems on the interval, J. Phys. A: Math. Theor. {\bf 45} (2012), 195203.

\bibitem{TF2008}
P. A. Treharne and A. S. Fokas, The generalized Dirichlet to Neumann map for the KdV equation on the half-line,
{\it J. Nonlinear Sci.} {\bf 18} (2008), 191--217. 

\end{thebibliography}

\end{document}